\documentclass[preprint,12pt]{elsarticle}

\usepackage{graphicx}
\usepackage{amssymb}

\usepackage{graphicx}
\usepackage[utf8]{inputenc}
\usepackage{lmodern}
\usepackage[T1]{fontenc}
\usepackage{lscape}
\usepackage{amsmath}	
\usepackage{amssymb}
\usepackage{cases}
\usepackage{graphicx}
\usepackage{caption}
\usepackage{comment}

\usepackage{makecell}
\usepackage{multirow}
\usepackage{amsfonts}
\usepackage{float} 
\usepackage{diagbox}
\usepackage{fullpage}
\usepackage{url}
\usepackage{rotating}
\usepackage{eurosym}
\usepackage{wrapfig}
\usepackage[final]{pdfpages} 
\usepackage{epstopdf} 
\usepackage[a4paper]{geometry}
\usepackage{placeins}
\usepackage{float}
\usepackage{listings}
\usepackage{color, colortbl}
\usepackage{hyperref}
\usepackage{xcolor}
\usepackage{graphicx, caption, subcaption}
\usepackage{tabularx}
\usepackage[inkscapelatex=false]{svg}
\usepackage{algorithm}
\usepackage{algpseudocode}

\definecolor{codegreen}{rgb}{0,0.6,0}
\definecolor{codegray}{rgb}{0.5,0.5,0.5}
\definecolor{codepurple}{rgb}{0.58,0,0.82}
\definecolor{backcolour}{rgb}{0.95,0.95,0.92}

\lstdefinestyle{mystyle}{
    backgroundcolor=\color{backcolour},
    commentstyle=\color{codegreen},
    keywordstyle=\color{magenta},
    numberstyle=\tiny\color{codegray},
    stringstyle=\color{codepurple},
    basicstyle=\ttfamily\footnotesize,
    breakatwhitespace=false,
    breaklines=true,
    captionpos=b,
    keepspaces=true,
    numbers=left,
    numbersep=5pt,
    showspaces=false,
    showstringspaces=false,
    showtabs=false,
    tabsize=2
}
\usepackage{sidecap}
\hypersetup{
colorlinks,
citecolor=black,
filecolor=black,
linkcolor=black,
urlcolor=black
}

\usepackage{natbib}

\newcommand{\blue}{\textcolor{black}}
\newcommand{\rev}{\textcolor{black}}

\journal{xxxx}

\begin{document}

\begin{frontmatter}


\title{TorchDA: A \rev{Python} package for performing data assimilation with deep learning forward and transformation functions}



\author{Sibo Cheng$^{1,2*}$, Jinyang Min$^{3}$, Che Liu$^{3}$, Rossella Arcucci$^{3}$ }

\address{   \small $^{1}$ CEREA, \'{E}cole des Ponts and EDF R\&D, \^Ile-de-France, France\\
\small $^{2}$ Data Science Instituite, Department of Computing, Imperial College London, UK\\
        \small $^{3}$ Department of Earth Science \& 
        Engineering, Imperial College London, UK  \\
        \hspace{5mm}\\
         corresponding: sibo.cheng@enpc.fr
}

\begin{abstract}
Data assimilation techniques are often confronted with challenges handling complex high dimensional physical systems, because high precision simulation in complex high dimensional physical systems is computationally expensive and the exact observation functions that can be applied in these systems are difficult to obtain. It prompts growing interest in integrating deep learning models within data assimilation workflows, but current software packages for data assimilation cannot handle deep learning models inside. This study presents a novel Python package seamlessly combining data assimilation with deep neural networks to serve as models for state transition and observation functions. The package, named TorchDA, implements Kalman Filter, Ensemble Kalman Filter (EnKF), 3D Variational (3DVar), and 4D Variational (4DVar) algorithms, allowing flexible algorithm selection based on application requirements. Comprehensive experiments conducted on the Lorenz 63 and a two-dimensional shallow water system demonstrate significantly enhanced performance over standalone model predictions without assimilation. The shallow water analysis validates data assimilation capabilities mapping between different physical quantity spaces in either full space or reduced order space. Overall, this innovative software package enables flexible integration of deep learning representations within data assimilation, conferring a versatile tool to tackle complex high dimensional dynamical systems across scientific domains.\\

\newpage
\textbf{Program summary} 

Program Title: TorchDA 

Developer’s repository link: https://github.com/acse-jm122/torchda

Licensing provisions: GNU General Public License version 3

Programming language: Python3 

External routines/libraries: Pytorch. \\

\noindent Nature of problem: Deep learning has recently emerged as a potent tool for establishing data-driven predictive and observation functions within data assimilation workflows. Existing data assimilation tools like OpenDA and ADAO are not well-suited for handling predictive and observation models represented by deep neural networks. This gap necessitates the development of a comprehensive package that harmonizes deep learning and data assimilation.\\

\noindent Solution method: This project introduces TorchDA, a novel computational tool based on the PyTorch framework, addressing the challenges posed by predictive and observation functions represented by deep neural networks. It enables users to train their custom neural networks and effortlessly incorporate them into data assimilation processes. This integration facilitates the incorporation of real-time observational data in both full and reduced physical spaces.

\end{abstract}




\end{frontmatter}

\clearpage
\section*{Main Notations}
\footnotesize{
\begin{table*}[ht!]
    \centering
    \begin{tabularx}{\textwidth}{ p{3.5cm} X }
        \hline
        \textbf{Notation} & \textbf{Description} \\
        \hline
        \\
        & \textit{Data Assimilation Algorithms} \\
        \\
        $t, k$ & continue and discrete time point \\
        $\mathbf{x}_k$ & state vector of the system at time $k$ \\
        $\mathbf{y}_k$ & observation vector at time $k$ \\
        $\mathbf{X}_{[:,-1]}$  & the last state vector in the state matrix $\mathbf{X}$ \\
        $\mathcal{M}_k$ & state transformation function \\
        $\mathbf{H}_k$ & linearised observation function \\
        $\mathbf{B}_k$ & background error covariance matrix \\
        $\mathbf{R}_k$ & observation error covariance matrix  \\
        $\mathbf{P}_k$ & forecast error covariance matrix  \\
        $\mathbf{K}_k$ & Kalman Gain matrix \\
        $\mathbf{I}$ & identity matrix \\
        $i$ & index of ensemble members \\
        $\mathbf{x}^{(i)}_k$ & state vector estimate for the $i$-th ensemble member \\
        $\boldsymbol{\varepsilon}_k^{(i)}$ & stochastic perturbation of each ensemble member $i$\\
        $\mathbf{y}^{(i)}_k$ & observation estimates for the $i$-th member \\
        $\gamma$ & learning rate of the optimisation algorithm \\
        $\mathcal{H}$ & nonlinear observation function \\
        $\mathbf{\bar x}_k$ & central state estimate at time $k$\\
        $\mathbf{x}_b$ & background state estimate \\
        $\hat{\mathcal{H}}$ & observation operator in the reduced space \\
        $\mathcal{J}$ & objective function of variational data assimilation \\
        \\
        & \textit{Lorenz 63 Test Case} \\
        \\
        $\sigma$ & ratio of kinematic viscosity to thermal diffusivity\\
        $r$ & ratio of the Rayleigh number to the critical value \\
        $\beta$ & geometrical factor \\
        \\
        & \textit{Shallow Water Test Case} \\
        \\
        $u, v$ & horizontal and vertical velocity components  \\
        $h$ & height of the fluid \\
        $E_u/D_u$ & encoder/decoder of the horizontal component \\
        $E_h/D_h$ & encoder/decoder of the vertical component \\
        $\hat{\mathbf{x}}$ & latent state estimate \\
        $\hat{\mathbf{y}}$ & latent observation \\
        \hline
    \end{tabularx}
\end{table*}}

\section{Introduction}
Traditional numerical simulation methods often encounter challenges such as high computational costs, slow processing speeds, and noticeable long-term prediction errors when dealing with high dimensional physical fields~\cite{cho2017numerical,rasmussen2006gaussian,gyorfi2002distribution,shutyaev2019methods}. To mitigate the predictive errors arising from forecasting models, data assimilation algorithms are frequently employed~\cite{shutyaev2019methods,carrassi2018data,cheng2023machine,amendola2020data,peyron2021latent}. For example, data assimilation can contribute to reliable ocean climate prediction~\cite{fujii2019observing}, and prove valuable in nuclear engineering applications~\cite{siefman2019development, gong2023parameter,gong2020inverse}. Other examples involve using data assimilation technique to improve the estimation of the hydrological state and fluxes~\cite{camporese2022recent}, and using data assimilation in fluid dynamics, especially for complex flows and turbulent phenomena~\cite{suzuki2015data}. Generally, data assimilation combines observational data with forecasting to enhance the fidelity of predictions and facilitate the dynamic updating of model states~\cite{shutyaev2019methods,carrassi2018data,cheng2023machine}. By iteratively refining model estimates through the integration of observations, data assimilation seeks to minimise the difference between predictions and observations~\cite{shutyaev2019methods,carrassi2018data,cheng2023machine}. However, the practical utilisation of data assimilation algorithms is based on the availability of explicit predictive and observation functions, thereby constraining their efficacy in scenarios characterised by challenges in function derivation, notably evident in instances of chaotic, high dimensional, or nonlinear systems~\cite{shutyaev2019methods,frerix2021variational,moradkhani2018fundamentals,halim2020deep}. Efforts to improve speed of simulations involve using data-driven methods, especially by using neural networks to approximate forward functions, but it is nontrivial to seamlessly integrate these data-driven approaches with conventional data assimilation tools~\cite{chen2022novel,panda2021data,penny2022integrating}.

In the last decade, deep learning has emerged as a powerful tool for creating predictions and observation models, especially when explicit functions are difficult to obtain. This advancement enables the establishment of data-driven predictive and observation functions by leveraging available state variables and observations~\cite{cheng2023machine}. Researchers have explored the integration of this innovative technique within their data assimilation workflows~\cite{cheng2023machine}. In a notable illustration, Amendola et al.~\cite{amendola2020data} harnessed autoencoders to encode carbon dioxide concentration states within a controlled environment into compact latent space vectors. Subsequently, they employed Long-Short Term Memory (LSTM) model to characterise the transitions within these encoded latent space representations~\cite{amendola2020data}. Correspondingly, Arcucci et al.~\cite{arcucci2021deep} and Boudier et al.~\cite{boudier2023data} reported noteworthy reductions in forecasting errors upon integrating deep learning with data assimilation, particularly within the context of the Lorenz system. Moreover, Cheng et al.~\cite{cheng2023generalised} utilised Generalised Latent Assimilation (GLA) technique and conducted experiments pertaining to flow in a tube, leading to substantial acceleration and remarkable reductions in computational expenses for predictive tasks over time. Another method named Latent Space Data Assimilation (LSDA) adopted Multi-Layer Perceptron (MLP) as the surrogate function to connect latent spaces directly, which leads to reduced computational costs~\cite{mohd2022deep}. In addition to improving prediction results by combining deep learning models and data assimilation algorithms, Barjard et al.~\cite{brajard2020combining} and Bocquet et al.~\cite{bocquet2020bayesian} also utilised deep learning and data assimilation combined approach to iteratively optimise the deep learning model, which was proved that deep learning models can achieve competitive performance on noisy and sparse observations with the assistance of data assimilation. However, all these instances necessitated the assembly of data assimilation workflows involving the integration of deep learning models. The work of Frerix et al.~\cite{frerix2021variational} provides an illustrative case, wherein they have made their code for training deep learning based inverse observation operators publicly accessible, but their workflow remains tied to their specific application. Hence, the development of a comprehensive package capable of seamlessly harmonising deep learning and data assimilation emerges as a necessary pursuit for the contemporary research community.

\rev{There are several data assimilation libraries and software tools which have been developed to facilitate the implementation of data assimilation algorithms. Noteworthy examples include OpenDA, DAPPER, PDAF and ADAO~\cite{openda-association-2023, raanes2024dapper, nerger2005pdaf,unknown-author-2023}.} OpenDA presents a convoluted and intricate configuration workflow, demanding a profound understanding of the software~\cite{openda-association-2023}. ADAO lacks seamless integration with modern programming environments, which impedes automatic code completion~\cite{unknown-author-2023}. \rev{DAPPER~\cite{raanes2024dapper} is an open-access Python package that benchmarks the results of various data assimilation methods and provides comparative studies.} However, the vital important defect is that these data assimilation tools are primarily suited for explicit predictive and observation functions in the context of background states assimilation~\cite{openda-association-2023,unknown-author-2023}. Therefore, upon an examination of well-established data assimilation tools such as OpenDA and ADAO, it becomes evident that existing data assimilation tools are insufficient for handling the predictive and observation models represented by deep neural networks.

\begin{figure}
     \centering
     \begin{subfigure}[c]{0.495\textwidth}
         \centering
         \includegraphics[width=\textwidth]{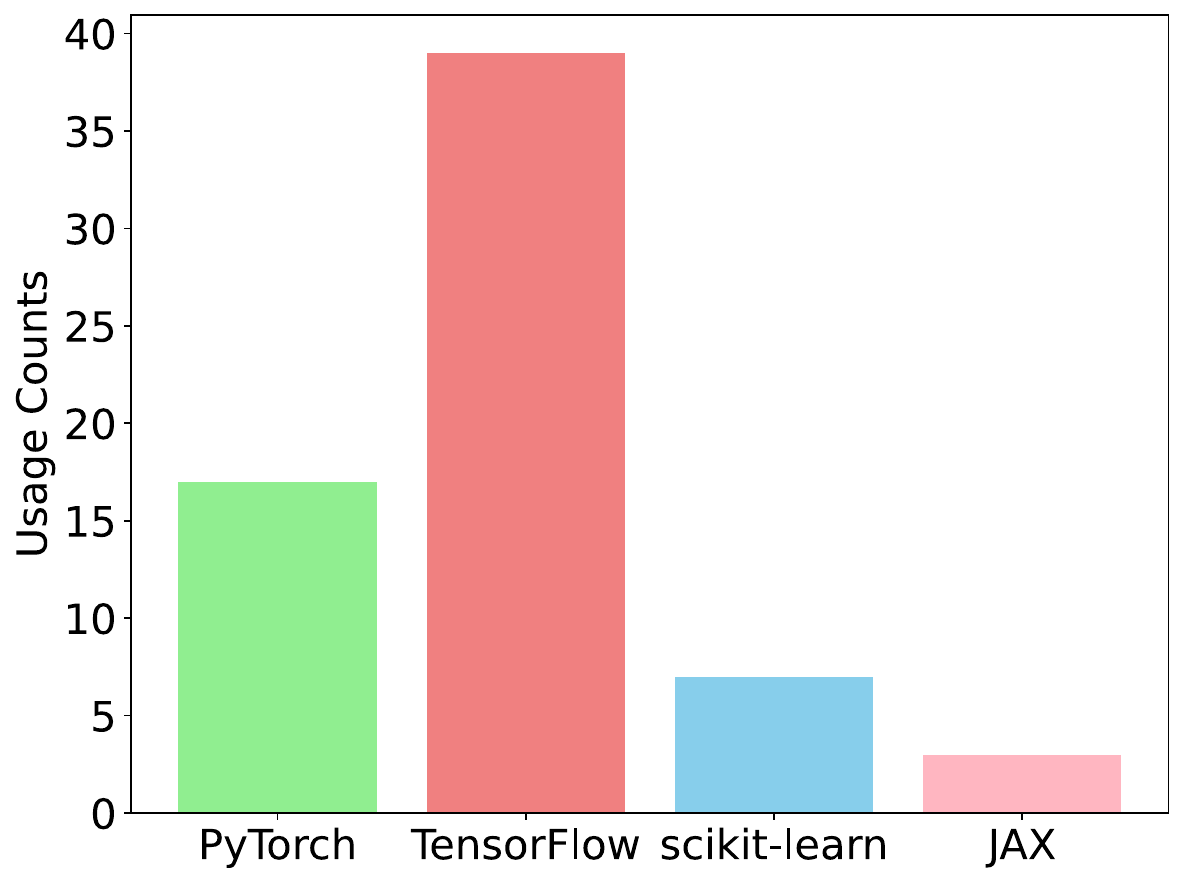}
         \caption{\label{fig:usage}Deep Learning Framework Usage Counts for Deep Learning and Data Assimilation Combined Applications}
     \end{subfigure}
     \begin{subfigure}[c]{0.495\textwidth}
         \centering
         \includegraphics[width=\textwidth]{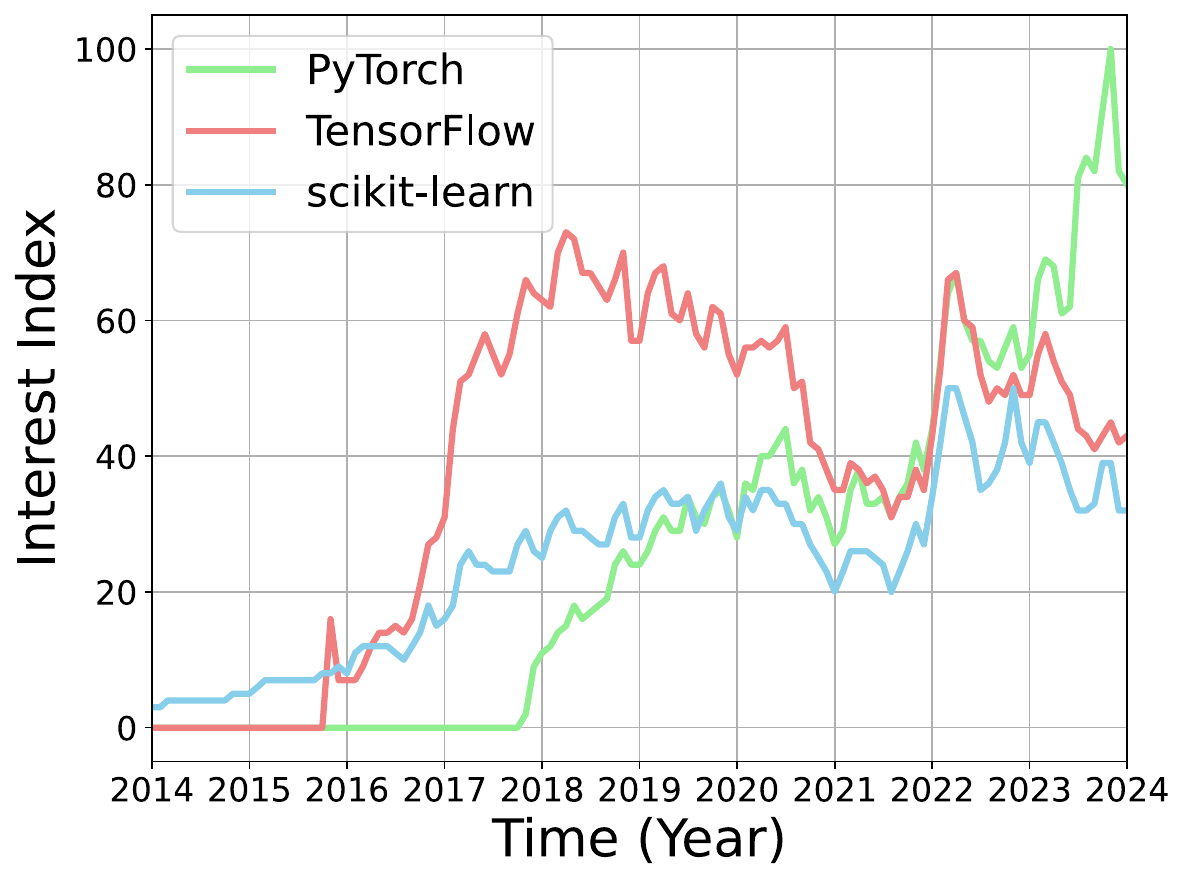}
         \caption{\label{fig:trends}Searching Trends of Different Deep Learning Frameworks on Google}
     \end{subfigure}
        \caption{Statistical Data of Deep Learning and Data Assimilation Combined Applications, and Interests Trend of Popular Machine Learning Frameworks on Google}
        \label{fig:usage and trends}
\end{figure}

The initial step in constructing a package that can integrate deep learning and data assimilation involves the selection of an appropriate deep learning framework capable of handling differentiation on neural networks. The statistical data on the usage of deep learning frameworks combined with data assimilation applications is presented in Figure~\ref{fig:usage} based on data assimilation related publications on \textbf{Papers With Code}\footnote{Papers With Code: \url{https://paperswithcode.com/}}. It is noticeable that TensorFlow is the most popular framework for the current data assimilation research community. However, an examination of Google search trends\footnote{Google Trends: \url{https://trends.google.com/trends/explore?date=2014-01-01\%202024-01-01\&q=\%2Fm\%2F0h97pvq,\%2Fg\%2F11bwp1s2k3,\%2Fg\%2F11gd3905v1\&hl=en-US}} for the general research community of Artificial Intelligence, as illustrated in Figure~\ref{fig:trends}, indicates that PyTorch is experiencing fluctuating growth trends rather than a gradual decline when compared to TensorFlow. Therefore, it is trustworthy that selecting PyTorch as the basis of this package for integrating deep learning and data assimilation could be the better choice for more potential studies of the data assimilation domain in the future.

In this project, a novel approach is proposed by integrating data assimilation with deep learning, and an innovative computational tool, named TorchDA, has been developed based on the PyTorch framework. This tool effectively addresses the data assimilation challenge posed by predictive and observation functions represented by deep neural networks. Unlike previously mentioned GLA and LSDA, utilising TorchDA eliminates the requirement for creating surrogate functions to connect distinct latent spaces. Users have the autonomy to train their own neural networks to replace conventional predictive and observation models. Subsequently, these neural networks can be integrated into TorchDA to facilitate data assimilation computations. \rev{In addition, implementing data assimilation algorithms using Pytorch will naturally allow the acceleration of GPU for parallel computing which is proved to be significantly more efficient compared to CPU computations~\cite{wei2023enable}. On the other hand, the stochastic gradient descent method can also improve the convergence of the assimilation problem by avoiding getting stuck in local minima~\cite{zhan2022efficient}.}

To verify the correctness of implementations inside this package, experiments on two testing cases including Lorenz 63 system~\cite{lorenz1963deterministic} and shallow water model~\cite{venant71Theorie} were conducted. These test cases involved the application of various data assimilation algorithms, including the Kalman Filter, Ensemble Kalman Filter (EnKF), 3D Variational (3DVar), and 4D Variational (4DVar). Specifically, the filter-based algorithms applied data assimilation on future prediction states over time, while variational methods concentrated on optimising the initial background state.

Compared to traditional data assimilation techniques driven by numerical models, this Python-based tool harnesses the capability of deep learning algorithms, markedly enhancing the efficiency of data assimilation computations. Moreover, this work successfully resolves the limitations existed in traditional Python data assimilation tools such as ADAO, particularly the incapability to handle derivatives of deep neural networks. \rev{In summary the main contributions of the proposed TorchDA package in the data assimilation community could be summarized as: }

\rev{\begin{itemize}
    \item A seamless integration of deep learning-based transformation and forward operator in data assimilation schemes;
    \item Embedded functionality for computing the tangent linear and adjoint models in deep learning frameworks;
    \item Efficient GPU computations for both Kalman-type and variational data assimilation algorithms;
    \item An easy implementation of stochastic gradient descent method to improve the DA optimization problem, in particular, for variational methods.
\end{itemize}}

The rest of this paper is organised as follows: it starts by presenting fundamental concepts of data assimilation and its combination with deep neural networks in Section~\ref{sec:back}, followed by the explanation of the package structure, then computational examples of Lorenz 63 Model and Shallow Water Model would be provided afterwards.

\section{Background}
\label{sec:back}
\subsection{Data Assimilation Concept}
As previously elucidated, data assimilation integrates observations with predictions of the state space, aiming to approximate the authentic state at a specific temporal instant~\cite{shutyaev2019methods,carrassi2018data,cheng2023machine}. A range of algorithms, notably including Kalman Filter-based techniques and variational methods, are commonly employed for data assimilation purposes~\cite{shutyaev2019methods,carrassi2018data}. Moreover, these methods often combine with deep neural networks in current studies, which indicated the importance of explaining these approaches in detail~\cite{cheng2023machine}.

\subsubsection{Kalman Filter}
The Kalman Filter is a fundamental algorithm in data assimilation because it is efficient to combine observed data and model predictions to yield refined state estimates~\cite{jazwinski2007stochastic, bain2009fundamentals}.

At its core, the Kalman Filter embodies an iterative process comprising two primary steps: the prediction step (Time Update) and the update step (Measurement Update)~\cite{jazwinski2007stochastic, bain2009fundamentals}. Let \(\mathbf{x}_k\) represents the state vector of the system at time \(k\), while \(\mathbf{y}_k\) signifies the corresponding observation vector. The prediction step engenders an estimate of the state vector \(\mathbf{x}_{k|k-1}\) based on the previous state \(\mathbf{x}_{k-1}\) \rev{and the linearised state transition \(\mathcal{M}_k\)}, while accounting for the perturbations introduced by the process noise \(\mathbf{Q}_k\):

\[
\mathbf{x}_{k|k-1} = \mathcal{M}_k \mathbf{x}_{k-1|k-1}.
\]

Simultaneously, the error covariance matrix \(\mathbf{P}_{k|k-1}\) evolves via the relation:

\[
\mathbf{P}_{k|k-1} = \mathcal{M}_k \mathbf{P}_{k-1|k-1} \mathcal{M}_k^T + \mathbf{Q}_k,
\]
where $(.)^T$ is the transpose operator.
Subsequently, the update step refines the state estimate and the associated error covariance matrix by assimilating the observational information via linearised observation function \(\mathbf{H}_k\). \rev{The Kalman Gain \(\mathbf{K}_k\) is calculated according to the error covariance matrix of the observations and predicted state vector.} The updated state estimate \(\mathbf{x}_{k|k}\) and the corresponding error covariance matrix \(\mathbf{P}_{k|k}\) is calculated as:

\[
\mathbf{K}_k = \mathbf{P}_{k|k-1} \mathbf{H}_k^T (\mathbf{H}_k \mathbf{P}_{k|k-1} \mathbf{H}_k^T + \mathbf{R}_k)^{-1},
\]
\[
\mathbf{x}_{k|k} = \mathbf{x}_{k|k-1} + \mathbf{K}_k (\mathbf{y}_k - \mathbf{H}_k \mathbf{x}_{k|k-1}),
\]
\[
\mathbf{P}_{k|k} = (\mathbf{I} - \mathbf{K}_k \mathbf{H}_k) \mathbf{P}_{k|k-1}.
\]

This process shows the ability of Kalman Filter to obtain better estimates of the current state by considering uncertainties in the system dynamics and measurement process~\cite{jazwinski2007stochastic, bain2009fundamentals}. However, the Kalman Filter can only tackle linear problems due to the requirement of transpose on both state transition \(\mathcal{M}_k\) and observation function \(\mathbf{H}_k\). The Kalman Filter implementation inside the package is described in Algorithm~\ref{algo:1}. \rev{It is worth mentioning that the Kalman Filter
implementation in TorchDA assumes a constant $\mathbf{P}$ matrix which is a simplification to a standard Kalman Filter.} In this context, the symbol $[\emptyset]$ denotes an empty list, and the operator $\oplus$ indicates the concatenation operation. In addition, $\mathbf{X}_{[:,-1]}$ refers to selecting the last state vector in the $\mathbf{X}_{estimates}$ matrix which is a sequence of state vectors along timeline.

\begin{algorithm}
\caption{Formulation of the implemented Kalman Filter}
Inputs: $\mathcal{M}, \mathbf{H}, \mathbf{P}, \mathbf{R}, \mathbf{x}_b,\mathbf{y}$
\begin{algorithmic}
\State parameters: $\mathbf{x}_k$, $\mathbf{X}_{estimates}$, $\rev{N_y}$
\State $\mathbf{x}_k = \mathbf{x}_b$, $\mathbf{X}_{estimates} \gets [\emptyset]$, $\rev{N_y} \gets \rev{\text{number of uncorrelated observations}}$
\For{$k \gets 1$ to $\rev{N_y}$}
    \State \rev{$\mathbf{X}_k = \mathcal{M}(\mathbf{x}_{k-1,estimates})$}
    \State $\mathbf{x}_k \gets \mathbf{X}_{k[:,-1]}$
    \State \rev{$\mathbf{x}_{k,estimates} = \mathbf{x}_k + \mathbf{P}\mathbf{H}^T[\mathbf{H}\mathbf{P}\mathbf{H}^T + \mathbf{R}]^{-1}(\mathbf{y}_k - \mathbf{H}\mathbf{x}_k)$}
    \State $\mathbf{X}_{estimates} \gets \mathbf{X}_{estimates} \oplus \mathbf{X}_k$
\EndFor
\end{algorithmic}
Output: $\mathbf{X}_{estimates}$

\label{algo:1}
\end{algorithm}

\subsubsection{Ensemble Kalman Filter}
The Ensemble Kalman Filter (EnKF) is a method that uses a group of model states to achieve better estimates of the current state~\cite{katzfuss2016understanding}. It supports nonlinear state transition \(\mathcal{M}_k\)~\cite{katzfuss2016understanding}. \rev{Meanwhile, the covariance matrix \(\mathbf{P}_k\) can be efficiently approximated by the ensemble process, so computational costs are also reduced~\cite{katzfuss2016understanding} compared to explicitly evolving the $\mathbf{P}$ matrix as done in a standard Kalman Filter. At the same time, the TorchDA package still allows for flow-dependent uncertainty estimates.}

In essence, the EnKF algorithm mirrors the principles of the Kalman Filter, albeit with a distinct ensemble-based approach~\cite{katzfuss2016understanding}. It operates iteratively, comprising a forecast step and an update step~\cite{katzfuss2016understanding}. At the heart of the EnKF lies the ensemble of model state vectors \(\{\mathbf{x}_k^{(i)}\}\), where \(i\) indexes the ensemble members, \(k\) signifies the time step, and \(\mathbf{x}_k^{(i)}\) denotes the state vector estimate for the \(i\)-th ensemble member at time \(k\).
\rev{In the current version of TorchDA, we consider a stochastic EnKF implementation~\cite{van2020consistent}.}

The forecast step engenders an ensemble of forecasts \rev{that is served as a proxy for the model predictions.} This step involves the perturbation of each ensemble member with a stochastic representation of the model's error, denoted by \(\boldsymbol{\varepsilon}_k^{(i)}\):

\[
\mathbf{x}_{k|k-1}^{(i)} = \mathcal{M}_k(\mathbf{x}_{k-1|k-1}^{(i)} + \boldsymbol{\varepsilon}_k^{(i)}).
\]

Subsequently, the update step refines the ensemble estimates by assimilating observational data denoted as \(\{\mathbf{y}_k^{(i)}\}\). This assimilation is pivotal as it introduces perturbations to the observations, thereby preserving the effectiveness of the ensemble process~\cite{burgers1998analysis}. This perturbation of observations serves to prevent the rapid decay of the entire variance of the ensemble, consequently preserving the capacity of ensemble to capture and represent the underlying system dynamics~\cite{burgers1998analysis}. The ensemble mean \(\bar{\mathbf{x}}_{k|k-1}\) is employed as a central estimate, and the covariance matrix \(\mathbf{P}_{k|k-1}\) quantifies the spread of the ensemble forecasts. The Kalman Gain \(\mathbf{K}_k\) is calculated based on the ensemble perturbations, and the updated ensemble estimates \(\mathbf{x}_{k|k}^{(i)}\) are computed as:

\[
\mathbf{K}_k = \mathbf{P}_{k|k-1} \mathbf{H}_k^T (\mathbf{H}_k \mathbf{P}_{k|k-1} \mathbf{H}_k^T + \mathbf{R}_k)^{-1},
\]
\[
\mathbf{x}_{k|k}^{(i)} = \mathbf{x}_{k|k-1}^{(i)} + \mathbf{K}_k (\mathbf{y}_k^{(i)} - \mathbf{H}_k \mathbf{x}_{k|k-1}^{(i)}).
\]

The procedure of EnKF is also presented in Figure~\ref{fig:enkf} for reference, and the implementation of EnKF inside the package is described in Algorithm~\ref{algo:2}.

\begin{algorithm}
\caption{Formulation of the implemented EnKF}
Inputs: $N_e, \mathcal{M}, \mathbf{H}, \mathbf{P}, \mathbf{R}, \mathbf{x}_b,\mathbf{y}$
\begin{algorithmic}
\State parameters: $\mathbf{X}_k$, $\mathbf{X}_{average}$, $\mathbf{X}_{ensemble}$, \rev{$\mathbf{X}_{estimates}$}, $\rev{N_y}$
\State $\mathbf{X}_k \gets N_e\ \text{number\ of\ ensemble\ perturbed\ around}\ \mathbf{x}_b\ \text{with}\ \mathbf{P}$
\State $\mathbf{X}_{average} \gets [\emptyset]$, $\rev{N_y} \gets \rev{\text{number of uncorrelated observations}}$
\For{$k \gets 1$ to $\rev{N_y}$}
    \State $\mathbf{X}_{mean} = 0$
    \For{$i \gets 1$ to $N_e$}
        \State \rev{$\mathbf{X}^{(i)} = \mathcal{M}(\mathbf{X}^{(i)}_{k-1,estimates})$}
        \State $\mathbf{X}^{(i)}_{k} \gets \mathbf{X}^{(i)}_{[:,-1]}$
        \State $\mathbf{X}_{mean} = \mathbf{X}_{mean} + \mathbf{X}^{(i)}$
    \EndFor
    \State $\mathbf{Y} \gets N_e\ \text{number\ of\ ensemble\ perturbed\ around}\ \mathbf{y}_k\ \text{with}\ \mathbf{R}$
    \State $\mathbf{x}_{mean} = \frac{1}{N_e} \sum_{e=1}^{N_e} \mathbf{X}_{k(e)}$, \rev{\text{where\ $\mathbf{X}_{k(e)}$\ is\ the\ e-th\ ensemble\ member}}
    \State $\mathbf{P}_e = \frac{1}{N_e-1} \cdot (\mathbf{X}_k-\mathbf{x}_{mean})^T(\mathbf{X}_k-\mathbf{x}_{mean})$
    \State \rev{$\mathbf{X}_{k, estimates} = \mathbf{X}_k + (\mathbf{P}_e\mathbf{H}^T[\mathbf{H}\mathbf{P}_e\mathbf{H}^T + \mathbf{R}]^{-1}(\mathbf{Y}^T - \mathbf{H}\mathbf{X}_k^T))^T$}
    \State $\mathbf{X}_{average} \gets \mathbf{X}_{average} \oplus (\frac{1}{N_e} \cdot \mathbf{X}_{mean})$
\EndFor
\end{algorithmic}
Outputs: $\mathbf{X}_{average}$

\label{algo:2}
\end{algorithm}

The EnKF algorithm is an ensemble-based approach which provides it with the capability to explicitly capture and propagate the inherent uncertainty within the system dynamics and observational measurements~\cite{katzfuss2016understanding,burgers1998analysis}. Moreover, the ensemble nature facilitates the representation of nonlinear features in the underlying state distribution, thus causing the EnKF algorithm particularly suitable for scenarios characterised by complex and nonlinear dynamics~\cite{katzfuss2016understanding,burgers1998analysis}.

\subsubsection{3D Variational data assimilation}
The 3D Variational (3DVar) algorithm embodies a variational approach aimed at optimising the state estimate by assimilating observations while accounting for prior information and background uncertainties~\cite{rabier2003variational}.

Fundamentally, the 3DVar framework formulates the assimilation process as an optimisation problem, with the objective of minimising the mismatch between background state and observational data~\cite{rabier2003variational}. At its core, the algorithm seeks to determine the state estimate \(\mathbf{x}_0\) that optimally merges the background state with the observational data~\cite{rabier2003variational}. This behaviour is accomplished through the construction of an objective function that quantifies the difference between the observed data \(\mathbf{y}\) and the background state estimate \( \mathbf{x}_{\text{b}}\):

\[
J(\mathbf{x}_0) = \frac{1}{2} (\mathbf{x}_0 - \mathbf{x}_{\text{b}})^T \mathbf{B}^{-1}_0 (\mathbf{x}_0 - \mathbf{x}_{\text{b}}) + \frac{1}{2} (\mathbf{y}_0 - \mathcal{H}(\mathbf{x}_0))^T \mathbf{R}^{-1}_0 (\mathbf{y}_0 - \mathcal{H}(\mathbf{x}_0)),
\]

where \(\mathbf{B}_0\) and \(\mathbf{R}_0\) signify the background error covariance matrix and the observation error covariance matrix, respectively, and \(\mathcal{H}\) represents the observation operator. The observation operator \(\mathcal{H}\) can be a nonlinear mapping from the state space to observation space. The optimisation process entails seeking the state estimate that minimises the objective function, thus yielding an assimilated state \(\mathbf{x}_0\) that effectively blends model predictions and observations~\cite{rabier2003variational}. The 3DVar implementation inside the package is described in Algorithm~\ref{algo:3}.

The variational nature of the 3DVar algorithm presents several advantages, including the ability to naturally accommodate prior information, exploit statistical properties of the background state, and incorporate observation uncertainties~\cite{rabier2003variational}. However, 3DVar only considers background state and observations at a single time point, which is solving a static problem by directly optimising a single estimate without regarding future evolutions.

\begin{algorithm}
\caption{Formulation and minimisation of the implemented 3DVar}
Inputs: $\mathcal{H}, \mathbf{B}, \mathbf{R}, \mathbf{x}_b,\mathbf{y}, N_{iterations}, \gamma$
\begin{algorithmic}
\State parameters: $\mathbf{x}_{b}^a$
\State $\mathbf{x}_{b}^a \gets \text{optimizable state vector copied from }\mathbf{x}_b$
\For{$n \gets 1$ to $N_{iterations}$}
    \State $\mathcal{J}(\mathbf{x}_b^a) = \vert \vert\mathbf{x}_b^a-\textbf{x}_{b}\vert \vert^2_{\textbf{B}^{-1}} + \vert \vert\textbf{y}-\mathcal{H}(\mathbf{x}_b^a)\vert \vert^2_{\textbf{R}^{-1}}$
    \State $\mathbf{x}_b^a \gets Adam(\mathcal{J}, \mathbf{x}_b^a, \gamma)$
\EndFor
\end{algorithmic}
Outputs: $\mathbf{x}_b^a$

\label{algo:3}
\end{algorithm}

\subsubsection{4D Variational data assimilation}
The 4D Variational (4DVar) algorithm is characterised by its intrinsic capability to assimilate a sequence of observations across a finite time interval, thus encompassing temporal dynamics and enhancing the accuracy of state estimation~\cite{rabier2003variational,talagrand20144d}.

As its 3DVar counterpart, the foundation of the 4DVar algorithm is positioned in variational principles. However, the 4DVar extends its scope beyond spatial dimensions to encompass the temporal dimension, thereby enabling the assimilation of observations within a predefined temporal window~\cite{rabier2003variational,talagrand20144d}. This temporal integration effectively provides the algorithm with the capability to address temporal evolution and capture intricate time-varying dynamics~\cite{rabier2003variational,talagrand20144d}. 
\begin{figure}
     \centering
     \begin{subfigure}[c]{0.476\textwidth}
         \centering
         \includegraphics[width=1\textwidth]{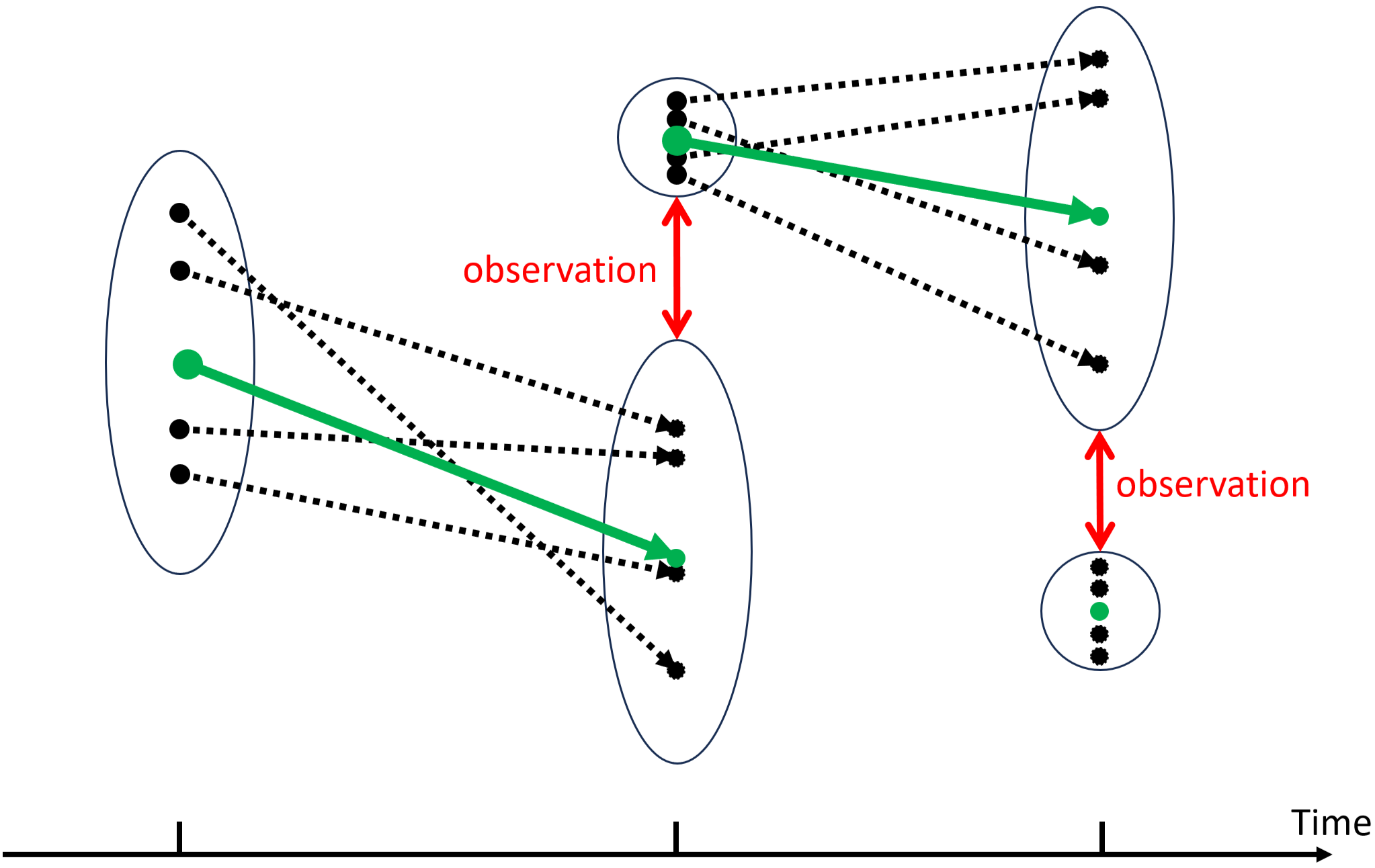}
         \caption{\label{fig:enkf}Ensemble Kalman Filter}
     \end{subfigure}
     \begin{subfigure}[c]{0.476\textwidth}
         \centering
         \includegraphics[width=1\textwidth]{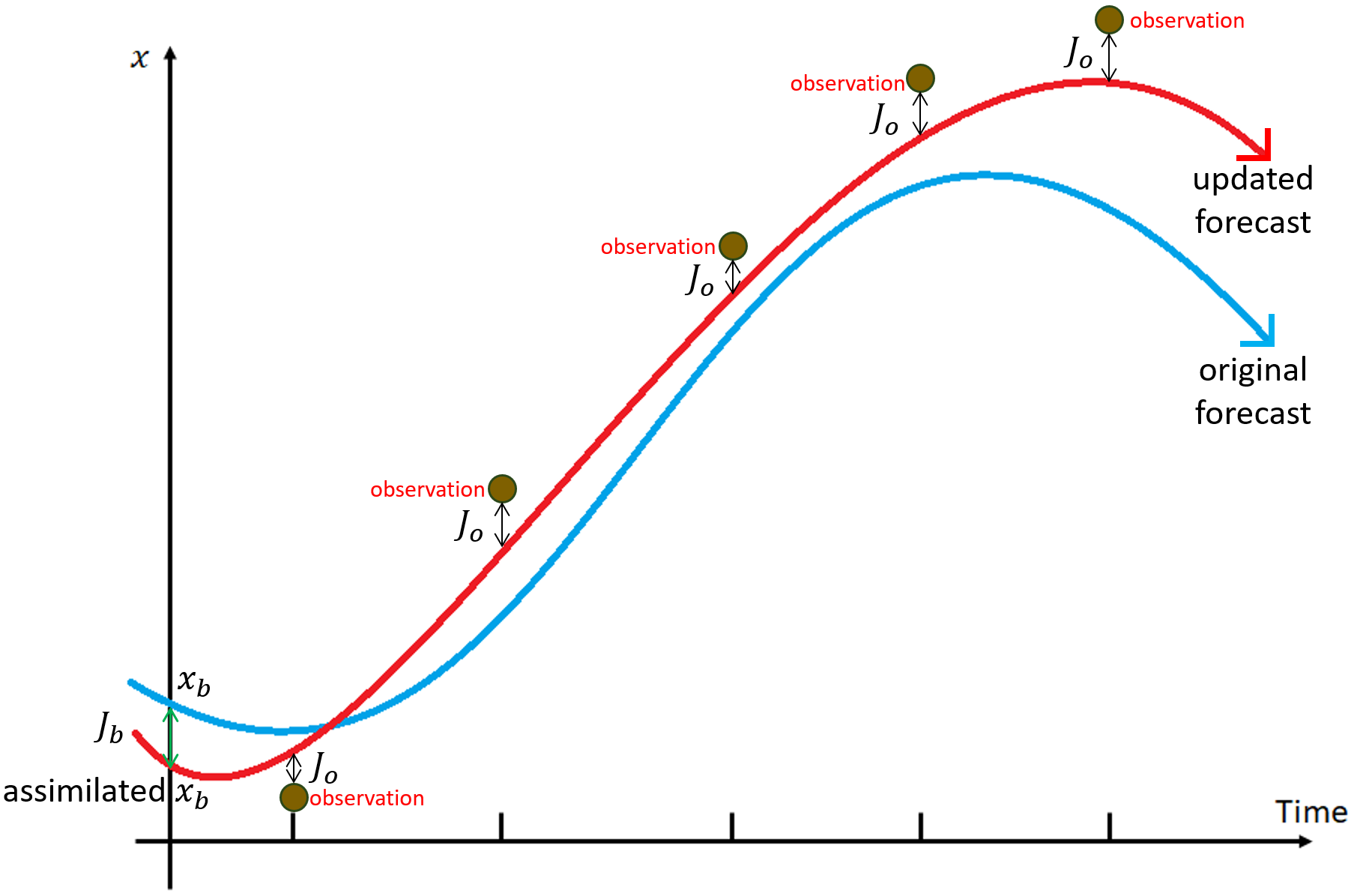}
         \caption{\label{fig:4dvar}4D Variational}
     \end{subfigure}
        \caption{Data Assimilation Algorithms}
        \label{fig:graphs of data assimilation algorithms}
\end{figure}
The 4DVar algorithm initiates an optimisation process aimed at ascertaining the optimal state trajectory that best aligns model predictions with observed data over the designated temporal interval~\cite{rabier2003variational,talagrand20144d}. The objective function represents a composite measure between the predicted states \(\mathcal{M}_k(\mathbf{x}_k)\) and the observed data \(\mathbf{y}_k\) at each assimilation time step, accounting for their temporal evolution:

\[
J(\mathbf{x}_0) = \frac{1}{2} (\mathbf{x}_0 - \mathbf{x}_{\text{b}})^T \mathbf{B}^{-1}_0 (\mathbf{x} - \mathbf{x}_{\text{b}}) + \frac{1}{2} \sum_{k=0}^{t} \left[ (\mathbf{y}_k - \mathcal{H} (\mathcal{M}_k(\mathbf{x}_k)))^T \mathbf{R}^{-1}_k (\mathbf{y}_k - \mathcal{H} (\mathcal{M}_k(\mathbf{x}_k))) \right],
\]

where the summation spans the assimilation time steps, and \(\mathbf{x}_{\text{b}}\) denotes the background state estimate for the initial time step. The optimisation process involves determining the state trajectory \(\mathbf{x}\) that minimises the aggregate objective function, yielding an assimilated state trajectory that captures both spatial and temporal dynamics~\cite{rabier2003variational,talagrand20144d}. The procedure of 4DVar is also presented in Figure~\ref{fig:4dvar} for reference, and implementation of 4DVar inside the package TorchDA is described in Algorithm~\ref{algo:4}.

The ability of the 4DVar algorithm to assimilate observations over a finite time window causes it particularly suited for scenarios characterised by rapidly evolving and time-dependent processes~\cite{rabier2003variational,talagrand20144d}. By integrating temporal information, the 4DVar technique contributes to a more reasonable true state estimation in the whole assimilation window than 3DVar, but 4DVar is more computational expensive due to the forward predictions required through the time window.

For more technical details about data assimilation algorithms, the readers are referred to the review papers~\cite{carrassi2018data} and~\cite{cheng2023machine}.

\begin{algorithm}
\caption{Formulation and minimisation of the implemented 4DVar}
Inputs: $\mathcal{M}, \mathcal{H}, \mathbf{B}, \mathbf{R}, \mathbf{x}_b,\mathbf{y}, N_{iterations}, \gamma$
\begin{algorithmic}
\State parameters: $\mathbf{x}_{b}^a$, $\rev{N_y}$
\State $\mathbf{x}_{b}^a \gets \text{optimizable state vector copied from }\mathbf{x}_b$
\State $\rev{N_y} \gets \rev{\text{number of uncorrelated observations}}$
\For{$n \gets 1$ to $N_{iterations}$}
    \State \rev{$\mathcal{J}_b(\mathbf{x}_b^a) = \vert \vert\mathbf{x}_b^a-\textbf{x}_{b}\vert \vert^2_{\textbf{B}^{-1}} $}
    \State $\mathbf{x} = \mathbf{x}_b^a$
    \State \rev{$\mathcal{J}_o = \vert \vert \textbf{y}_{(1)}-\mathcal{H}(\mathbf{x}_b^a)\vert \vert^2_{\textbf{R}^{-1}}$}
    \For{$k \gets 2$ to $\rev{N_y}$}
        \State $\mathbf{X} = \mathcal{M}(\mathbf{x})$
        \State $\mathbf{x} \gets \mathbf{X}_{[:,-1]}$
        \State $\mathcal{J}_o(\mathbf{x}) = \mathcal{J}_o(\mathbf{x}) + \vert \vert \textbf{y}_{(k)}-\mathcal{H}(\mathbf{x})\vert \vert^2_{\textbf{R}^{-1}}$
    \EndFor
    \State $\mathcal{J} = \mathcal{J}_b + \mathcal{J}_o$
    \State $\mathbf{x}_b^a \gets Adam(\mathcal{J}, \mathbf{x}_b^a, \gamma)$
\EndFor
\end{algorithmic}
Outputs: $\mathbf{x}_b^a$

\label{algo:4}
\end{algorithm}

\subsection{Integrating Data Assimilation and Deep Learning}
\begin{figure}
    \begin{center}
        \includegraphics[width=0.85\textwidth]{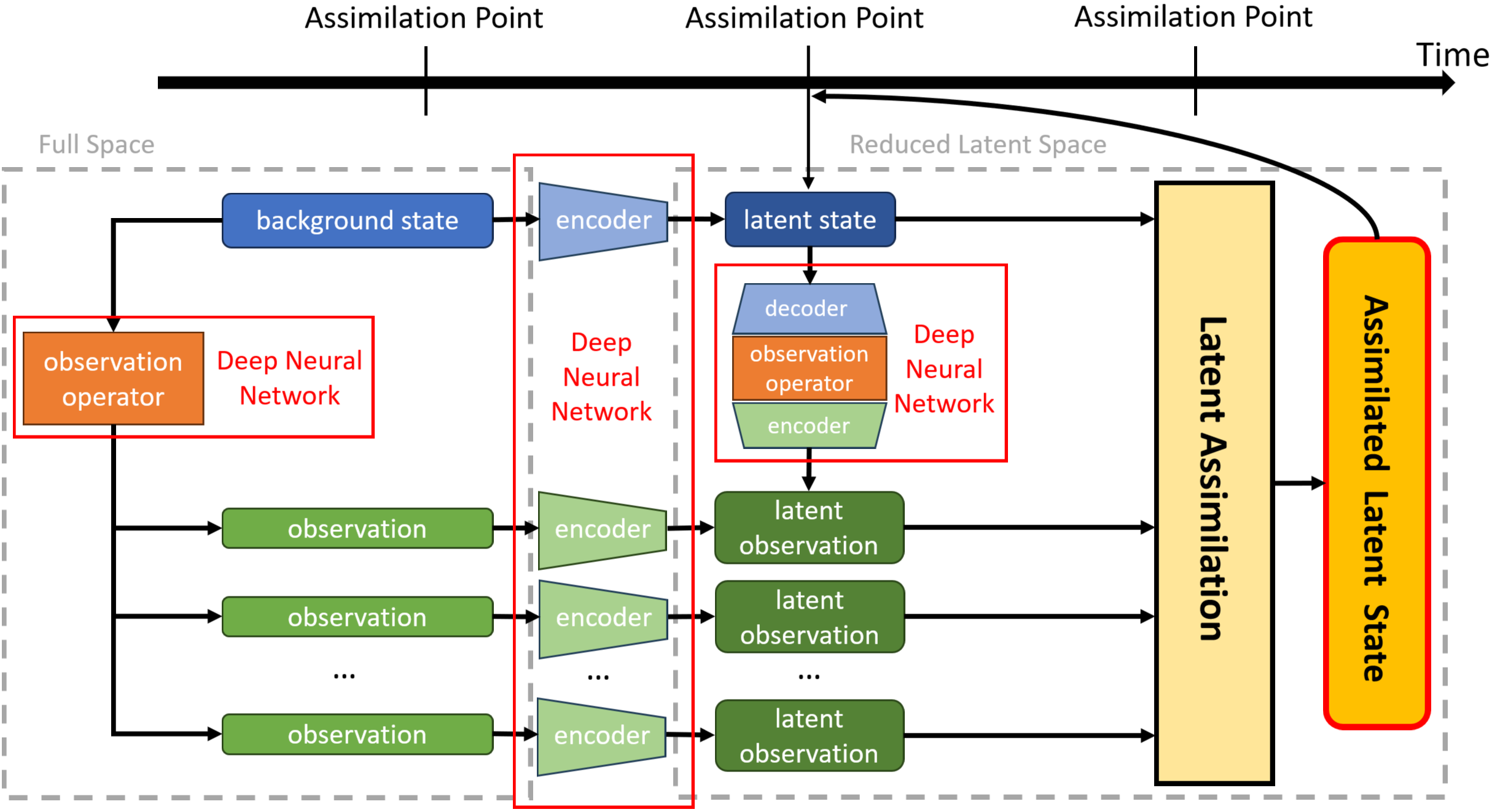}
    \end{center}
    \caption{\label{fig:latent_assimilation}Schematic Diagram of the Latent Data Assimilation with Deep Neural Network Surrogate Models}
\end{figure}

As previously demonstrated, there is a growing trend in various applications and experiments that aim to combine data assimilation with deep learning algorithms. Given the computational challenges entailed by full space data assimilation, a pragmatic approach involves conducting data assimilation within a reduced space~\cite{arcucci2019optimal,binev2017data,cheng2021observation,xiao2018parameterised}. Recent research endeavours have addressed the imperative of computational efficiency by integrating data assimilation with deep learning through the utilisation of autoencoders~\cite{wang2022deep,cheng2022data}. These algorithms, collectively referred to as latent assimilation, effectively harness the computational efficacy inherent in deep learning, while preserving the precision of data assimilation. Within a specific subset of latent assimilation methodologies, autoencoders are trained on individual physical quantities with the objective of reconstructing the original physical quantities within their respective representation spaces. The compressed representations situated at the bottleneck layer, positioned between the encoder and decoder components of the autoencoder, are subsequently employed for predictions and assimilation in latent space. To perform effective latent assimilation on state vector, these compressed representations are usually mapped into different latent spaces, because this process can involve more information from different physical quantities. The schematic depiction of this latent assimilation concept is illustrated in Figure~\ref{fig:latent_assimilation}.

Regarding latent assimilation, it falls within two broad methodological categories. The first involves the incorporation of latent state space data into full space observations during the assimilation process, while the second entails the utilisation of latent state space data with latent space observations. Peyron et al.~\cite{peyron2021latent} and Maulik et al.~\cite{maulik2022efficient} adopted a strategy wherein observations in the full physical space are employed to refine reduced order models, which aligns with the first category. Conversely, several other researchers have explored different approaches involving the compression of both state variables and observations into latent space~\cite{amendola2020data,cheng2022data,mack2020attention,liu2022enkf}. In addition to latent representation learned by autoencoder, Casas et al.~\cite{casas2020reduced} propose a methodology that employs Recurrent Neural Networks (RNNs) to acquire assimilated results within a reduced space produced by Principal Components Analysis (PCA), thereby enhancing future predictions. Therefore, the emergence of latent assimilation emphasises the necessity for a more streamlined integration of data assimilation and deep learning.

In addition to employing autoencoders for representation compression, recent research endeavours have also explored the substitution of the observation operator, which typically maps the state space to observation space, with deep neural networks. An illustrative study conducted by Frerix et al.~\cite{frerix2021variational} involved the training of a deep neural network to serve as an inverse observation operator, responsible for mapping the observation space back to the state space. This was accomplished through the modification of the cost function used during the optimisation steps, employing a variational approach within the domain of data assimilation~\cite{frerix2021variational}. Conversely, Storto et al.~\cite{storto2021neural} adopted a different approach by training a standard forward observation operator, which maps state space variables to observation space. However, they introduced a linearisation step for this operator around the background state~\cite{storto2021neural}. This circumvented the need for directly solving the cost function involving optimisation through a neural network. Consequently, it is evident that neural networks are increasingly being considered as potential substitutes for observation operators in the realm of data assimilation.

\section{Package Structure}
The software package is developed in Python and relies only on PyTorch as its unique dependency. The package encompasses five distinct modules, namely \textbf{parameters}, \textbf{builder}, \textbf{executor}, \textbf{kalman\_filter}, and \textbf{variational}. This Python package offers a versatile and user-friendly framework tailored for the seamless integration of data assimilation with neural networks across various algorithms, including Kalman Filter (KF), EnKF, 3DVar, and 4DVar. Below, a summary of the primary contents of each module is provided:

\textbf{parameters:} This module contains the Parameters class, enabling users to specify data assimilation parameters.

\textbf{builder:} The CaseBuilder class within this module facilitates the configuration and execution of data assimilation cases.

\textbf{executor:} The \_Executor class implemented here is responsible for executing data assimilation cases.

\textbf{kalman\_filter:} This module incorporates implementations of KF and EnKF algorithms for data assimilation.

\textbf{variational:} The variational module accommodates the 3DVar and 4DVar algorithms for variational data assimilation.

To understand more detailed prerequisites of this package imposed on users, it is recommended to read documents provided in the GitHub repository of this package \footnote{TorchDA: \url{https://github.com/acse-jm122/torchda}}. Some essential prerequisites of inputs and promise of outputs outlined as follows.

\textbf{Covariance Matrix $\mathbf{B}$ and $\mathbf{R}$:}

The software imposes specific constraints on the error covariance matrices B and R; both must be positive definite matrices with row and column dimensions aligned with the background state vector length or observational state vector length accordingly.

\textbf{Background State $\mathbf{x}_0$:}

The background state, denoted as $\mathbf{x}_0$, must conform to a 1D or 2D tensor with the shape ([$batch\ size$], state dimension), with $batch\ size$ being an optional parameter. $batch\ size$ is exclusively available for the 3DVar algorithm.

\textbf{Observations $\mathbf{y}$:}

Observations or measurements, referred to as $\mathbf{y}$, are required to be in the form of a 2D tensor. For the KF, EnKF, and 4DVar algorithms, the shape of $\mathbf{y}$ should be (\rev{the state dimension of observations}), with the number of observations being at least 1 in KF or EnKF, and a minimum of 2 in 4DVar. For the 3DVar algorithm, the shape of $\mathbf{y}$ must be ([$batch\ size$], state dimension), and $batch\ size$ is optional, with the default being $batch\ size$ equal to 1.

\textbf{State transformation function $\mathcal{M}_k$:}

The state transformation function, denoted as $\mathcal{M}_k$, should be a callable code object capable of handling a 1D tensor input $\mathbf{x}_0$ with the shape (state dimension,). The output of $\mathcal{M}_k$ must be a 2D tensor with the shape (time window sequence length, state dimension).

\textbf{Observation Operator $\mathbf{H}_k$ or $\mathcal{H}$:}

The observation operator, represented as $\mathbf{H}_k$ or $\mathcal{H}$, can take the form of either a tensor or a callable code object. In the KF algorithm, $\mathbf{H}_k$ can be a tensor. However, if $\mathcal{H}$ is a callable code object, it should be equipped to process a 2D tensor input $\mathbf{x}_0$ with the shape ([$number\ of\ ensemble$], state dimension). By default, the number of ensembles is set to 1 for all algorithms except EnKF. The output of the callable $\mathcal{H}$ code object should correspondingly possess the shape ([$number\ of\ ensemble$], measurement dimension), while in all other algorithms, the output should be a 2D tensor with the shape (1, state dimension).

In the case of KF, where a callable $\mathcal{H}$ code object is utilised, it is advisable to work with a 1D tensor input $\mathbf{x}_0$ (state dimension,) to mitigate potential issues arising from Jacobian approximations and output uncertainties in this process. While KF has not been included as an option in algorithm parameters due to these considerations, it is still accessible as an individual function for customisation in user applications.

\textbf{Output Structure:}

Execution results are returned in the form of packed native Python dictionaries. Two distinct sets of outputs are available for various underlying data assimilation algorithms:

1. In the context of the EnKF algorithm, it is possible to obtain two distinct categories of output: namely, "average\_ensemble\_all\_states" and "each\_ensemble\_all\_states". The former represents the average ensemble states across the entire forwarding time window in the EnKF process, while the latter portrays the result states of each ensemble member throughout the entire forwarding time window.

2. For the 3DVar or 4DVar algorithms, the output includes "assimilated\_state" and "intermediate\_results". The "assimilated\_state" signifies the optimised background state after the respective variational algorithm has completed its run. The "intermediate\_results" encompass a sub-level dictionary, containing "J", "J\_grad\_norm", and "background\_states" for each iteration in 3DVar, and "Jb", "Jo", "J", "J\_grad\_norm", and "background\_states" for each iteration in 4DVar.

To provide a practical demonstration of how to use this package, some actual code appeared in experiments on computational examples were shown in code snippets. Users have the option to input parameters using a basic dictionary, as exemplified in code snippet~\ref{lst:1}, but it is advisable to utilise an instance of Parameter class, as demonstrated in code snippet~\ref{lst:2}, which not only helps the use of code completion tools but also minimises the likelihood of typing errors by hands. However, the most recommended approach is to employ setters, as shown in code snippet~\ref{lst:3}. This method follows to the builder pattern, allowing users to dynamically configure parameters during runtime, which avoids the need to specify all parameters simultaneously. There are a total of 7 required parameters and 10 optional parameters depending on the specific assimilation algorithm choices. The required parameters consist of "algorithm," "device," "observation\_model," "background\_covariance\_matrix," "observation\_covariance\_matrix," "background\_state," and "observations". These parameters are crucial for executing the selected algorithm. For instance, users must specify the device they intend to use for running their applications, with current supported options including CPU and GPU. Additionally, if optimisation based algorithms include 3DVar and 4DVar are chosen, users need to assign values for learning rate, maximum iterations, and optimisation method. The current underlying optimisation algorithm employed is Adam~\cite{kingma2014adam}, where learning rate serves as its unique associated parameter.

\begin{lstlisting}[language=Python, caption=Passing Parameters by a Python Dictionary, label={lst:1}]
params_dict = {
    "algorithm": torchda.Algorithms.Var4D,
    "observation_model": H,
    "background_covariance_matrix": B,
    "observation_covariance_matrix": R,
    "background_state": xb,
    "observations": y,
    "forward_model": M,
    "output_sequence_length": gap + 1,
    "observation_time_steps": time_obs,
    "gaps": [gap] * (len(time_obs) - 1),
    "learning_rate": 7.5e-3,
    "args": (rayleigh, prandtl, b),
}
results_4dvar = torchda.CaseBuilder().set_parameters(params_dict).execute()
\end{lstlisting}

\begin{lstlisting}[language=Python, caption=Passing Parameters by a Parameter Object, label={lst:2}]
parameters = torchda.Parameters(
    algorithm=torchda.Algorithms.EnKF,
    device=torchda.Device.CPU,
    observation_time_steps=time_obs,
    gaps=gaps,
    num_ensembles=Ne,
    observation_model=H,
    output_sequence_length=lstm_model.out_seq_length,
    forward_model=lstm_model,
    background_covariance_matrix=P0,
    observation_covariance_matrix=R,
    background_state=xtT[0],
    observations=y.T,
)
run_case = torchda.CaseBuilder(parameters=parameters)
results = run_case.execute()
xEnKF = results["average_ensemble_all_states"]
x_ens = results["each_ensemble_all_states"]
\end{lstlisting}

\begin{lstlisting}[language=Python, caption=Passing Parameters by Setters, label={lst:3}]
case_to_run = (
    torchda.CaseBuilder()
    .set_observation_model(H)
    .set_background_covariance_matrix(B)
    .set_observation_covariance_matrix(R)
    .set_learning_rate(5)
    .set_max_iterations(300)
    .set_algorithm(torchda.Algorithms.Var3D)
    .set_device(torchda.Device.GPU)
)
### In the forward prediction loop ###
case_to_run.set_background_state(out).set_observations(y_imgs)
out = case_to_run.execute()["assimilated_state"]
\end{lstlisting}

\begin{figure}
    \begin{center}
        \includegraphics[width=0.65\textwidth]{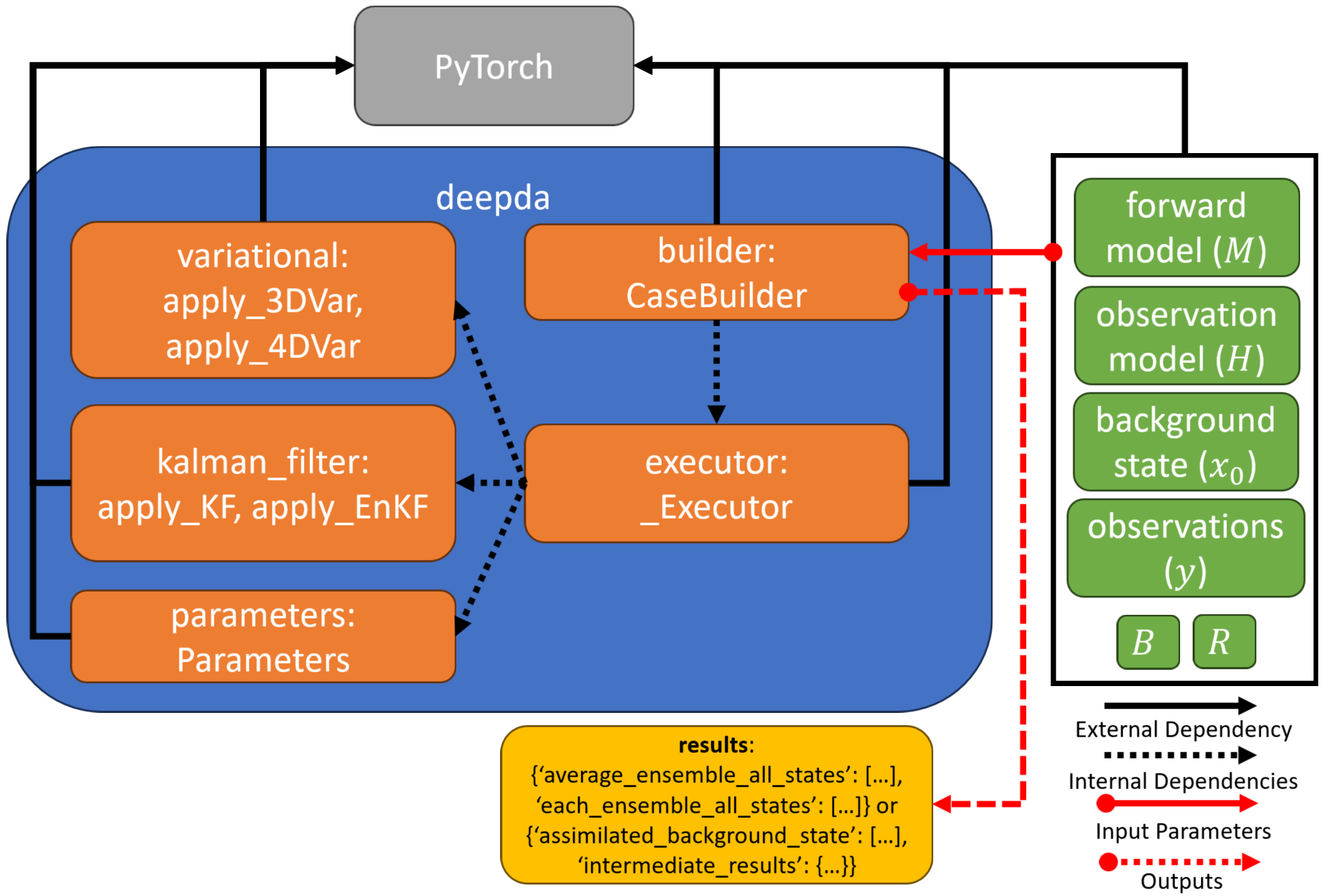}
    \end{center}
    \caption{\label{fig:diagram}Package Structure and Dependencies}
\end{figure}

This comprehensive description outlines the salient features and requirements of the software package, aiding users in effectively utilising it for data assimilation integrated with deep learning across a spectrum of algorithms. The comprehensive structure of the package arrangement is also visually presented in Figure~\ref{fig:diagram}.

\section{Computational Examples}
To validate the functionality and correctness of the package while further enhancing comprehension of data assimilation algorithms, comprehensive assessments were conducted. These assessments involved the application of deep neural networks, encompassing convolutional autoencoders, residual networks, and multilayer LSTM networks, to two distinct models: the Lorenz 63 model~\cite{lorenz1963deterministic}, and the shallow water model~\cite{venant71Theorie}. The rationale for selecting the Lorenz 63 model and the shallow water model as illustrative examples is rooted in their inherent chaotic and highly dynamic nature~\cite{lorenz1963deterministic,stewart2008correlated,cioaca2014low}, causing them to be ideal scenarios for the integration of deep neural networks as surrogate models within the context of data assimilation~\cite{stewart2008correlated,cioaca2014low}. For the Lorenz 63 model, the performance of EnKF and 4DVar were evaluated, while employing a multilayer LSTM network for forward predictions. In the case of the shallow water model, 3DVar and 4DVar were combined with autoencoders, residual networks, and multilayer LSTM networks separately. Specifically, autoencoders were employed to condense representations of full space physical quantities into reduced latent space representations, a residual network served as the observation operator, and a multilayer LSTM network was utilised for forward predictions in latent space.

\subsection{Lorenz 63 Model}
Lorenz 63 system provides a simplified yet chaotic representation of dynamic systems, facilitating comprehensive assessments of data assimilation algorithms~\cite{evensen2022enkf,evensen20223dvar}. Its inherent chaos and sensitivity to initial conditions, along with a known mathematical solution, establish an appropriate benchmark for evaluating the capabilities of algorithms~\cite{lorenz1963deterministic}. Specifically, the progression of the Lorenz 63 model is delineated through a set of differential equations presented below.

\begin{align}
&\frac{\mathrm{d}X}{\mathrm{d}t}=-\sigma(X-Y), \\
&\frac{\mathrm{d}Y}{\mathrm{d}t}=-XZ+rX-Y, \\
&\frac{\mathrm{d}Z}{\mathrm{d}t}=XY-\beta Z,
\end{align}

where $\{X, Y, Z\} \in \mathbb{R}^3$. $\sigma$ represents the Prandtl number, denoting the ratio of kinematic viscosity to thermal diffusivity~\cite{lorenz1963deterministic}. The parameter $r$ signifies the ratio of the Rayleigh number $R_a$ to the critical value of the Rayleigh number $R_{a_c}$, and $\beta$ represents a geometrical factor~\cite{lorenz1963deterministic}. 

\begin{figure}
    \begin{center}
        \includegraphics[width=\textwidth]{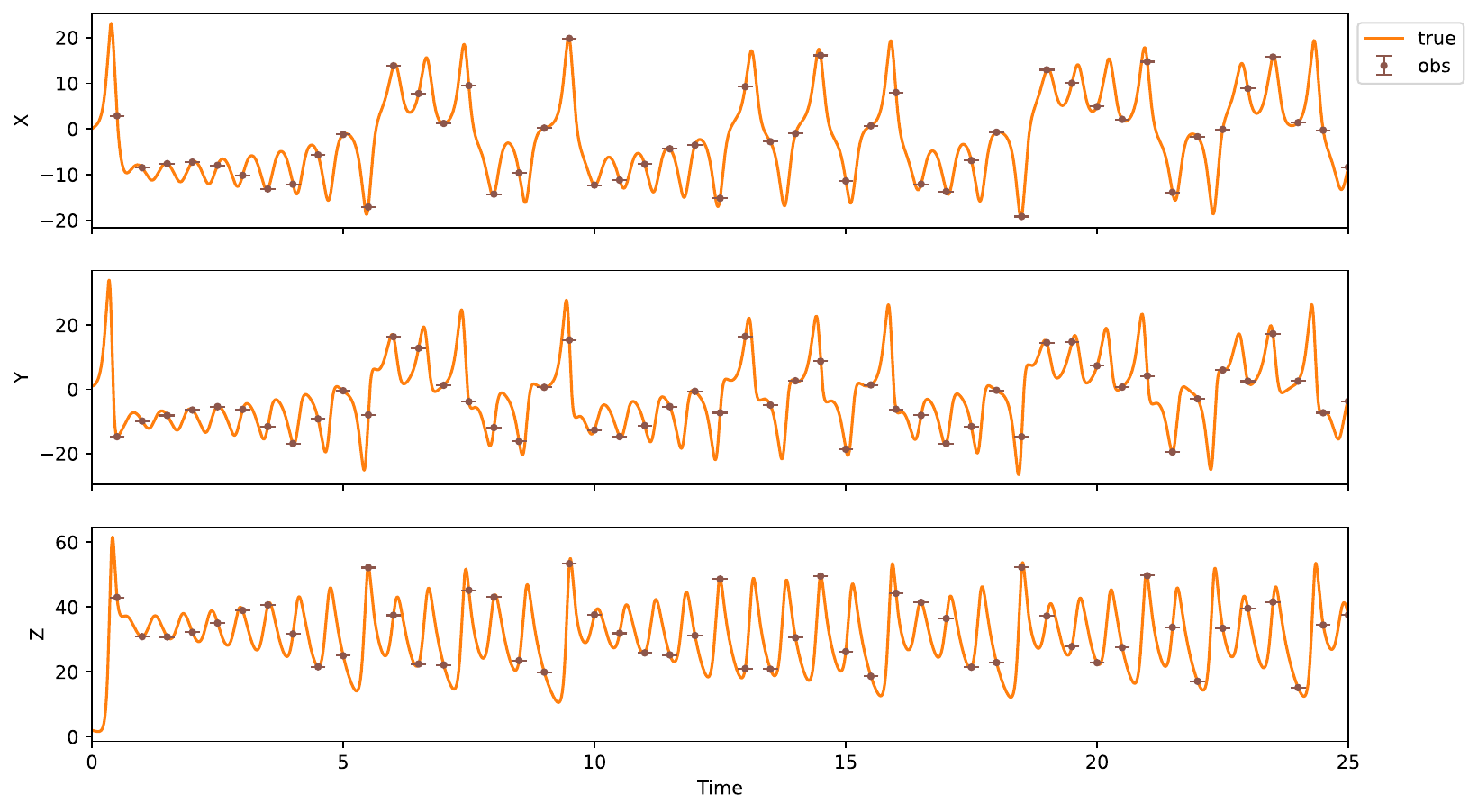}
    \end{center}
    \caption{\label{fig:reference_traj_and_obs}Reference Trajectory with Noisy Observations}
\end{figure}

Given the inherent uncertainty associated with the Lorenz 63 system, particularly regarding its evolving trajectory for different initial conditions, a standardised initial state has been established as $[X,Y,Z]^{T}=[0,1,2]^{T}$. The control parameters for the system have been fixed as follows: $\sigma=10$, $r=35$, and $\beta=8/3$. The temporal intervals between each state update have been set at $10^{-3}$ \blue{time units}, and the total evolution duration has been defined as $25$ \blue{time units}. The numerical integration technique employed for this evolution is the forward Euler method. Following the execution of this simulation, the resulting data is visually depicted in the Figure~\ref{fig:reference_traj_and_obs}.

To establish a forward model for sequential prediction, a 20-layer LSTM network was employed. This LSTM network possesses an input dimension of 3 and a hidden dimension of 12. It was trained with the parameter settings previously mentioned, except for the total evolution time, which was extended to $500$ \blue{time units} to ensure the sufficiency of training data. This network has the capability to predict 99 future time steps based on a given state vector at the current time step, because testing effects of data assimilation in long-term predictions is the purpose. To emulate this characteristic, the sampling frequency for observations was established at $0.5$ \blue{time units}. This implies that there are ${0.5}/{10^{-3}}=500$ state evolutions occurring between each observation point. Additionally, in realistic scenarios, instruments used for measurements are susceptible to introducing measurement noise. In this context, the standard deviation of the observation noise was configured to $10^{-3}$ to simulate the presence of a moderately accurate instrument during measurements. Figure~\ref{fig:reference_traj_and_obs} provides a representation of the outcomes resulting from this noisy sampling process.

\subsubsection{Experiment without Data Assimilation}
In the baseline setting, the multilayer LSTM network performed forward predictions without any corrections. Following the execution of predictions within a specified time window using only the forward model, an evaluation of errors on each component was conducted. This evaluation involved the utilisation of the Relative Root Mean Square Error (RRMSE) metric, which is the Mean Square Error (MSE) on each component normalised by the mean square value of the corresponding component in the reference trajectory. By employing this normalised relative error calculation, both the predictions generated by the forward model and the reference trajectory were analysed. The resulting plots in Figure~\ref{fig:reference_traj_and_forward_pred} depict the errors observed in each component relative to the reference trajectory.

It is worth noting that the trajectory predicted by the forward model exhibits a notable disparity when compared to the reference trajectory. This phenomenon can be attributed to the chaotic nature of the Lorenz 63 system~\cite{lorenz1963deterministic}. Chaotic systems like this exhibit unstable and nonperiodic behaviour, even when subjected to tiny numerical variations in initial conditions~\cite{lorenz1963deterministic}. Consequently, predicting the behaviour of this system accurately at each time step, even with the aid of a potent sequential model such as the 20-layer LSTM network employed in this instance, remains a significant challenge.

\subsubsection{Experiment of EnKF}
\begin{figure}
     \centering
     \begin{subfigure}[t]{1\textwidth}
         \centering
         \includegraphics[width=\textwidth]{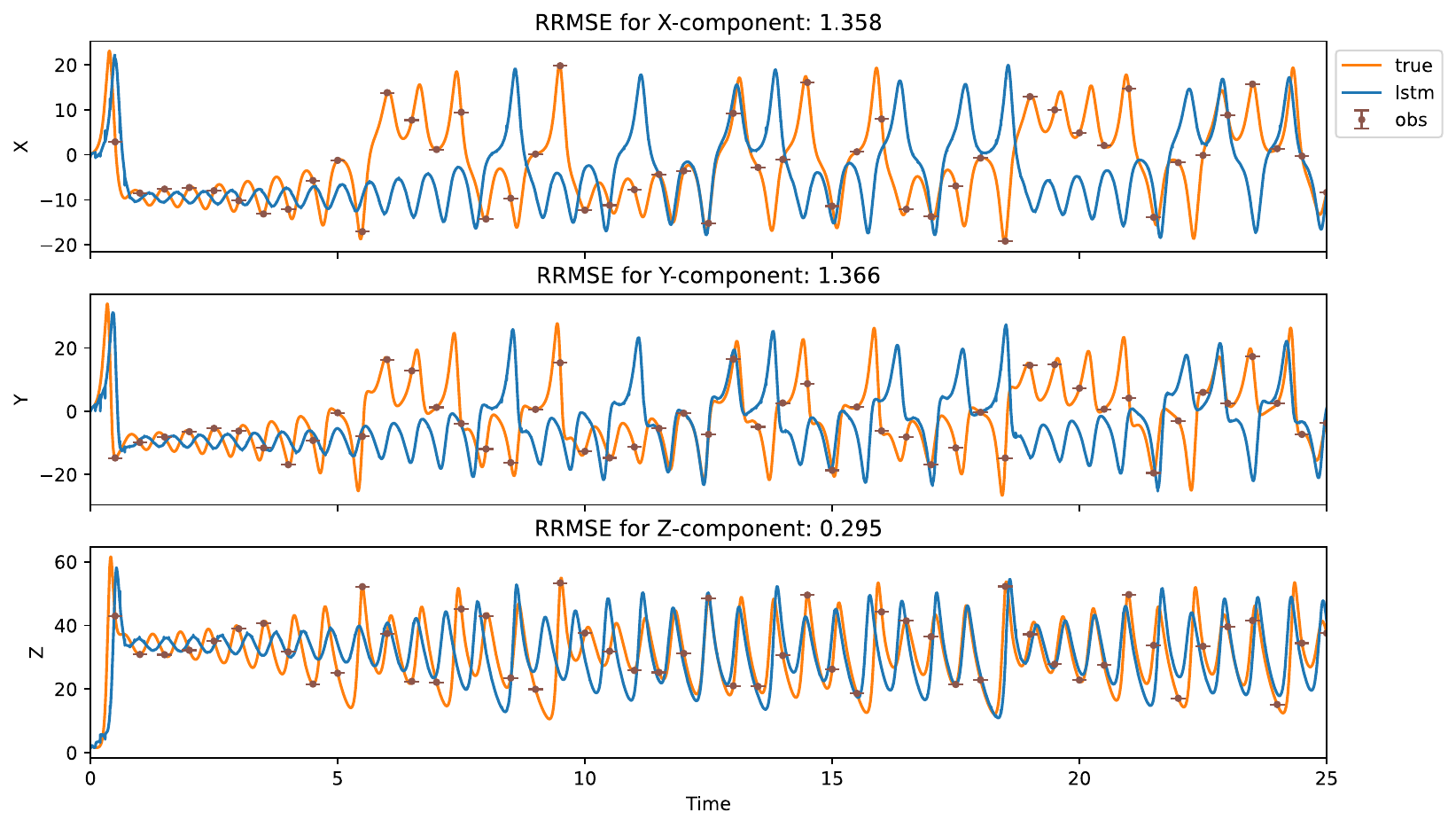}
         \caption{\label{fig:reference_traj_and_forward_pred}RRMSE on each Component between Reference Trajectory and Trajectory of Forward Model Prediction}
     \end{subfigure}
     \begin{subfigure}[t]{1\textwidth}
         \centering
         \includegraphics[width=\textwidth]{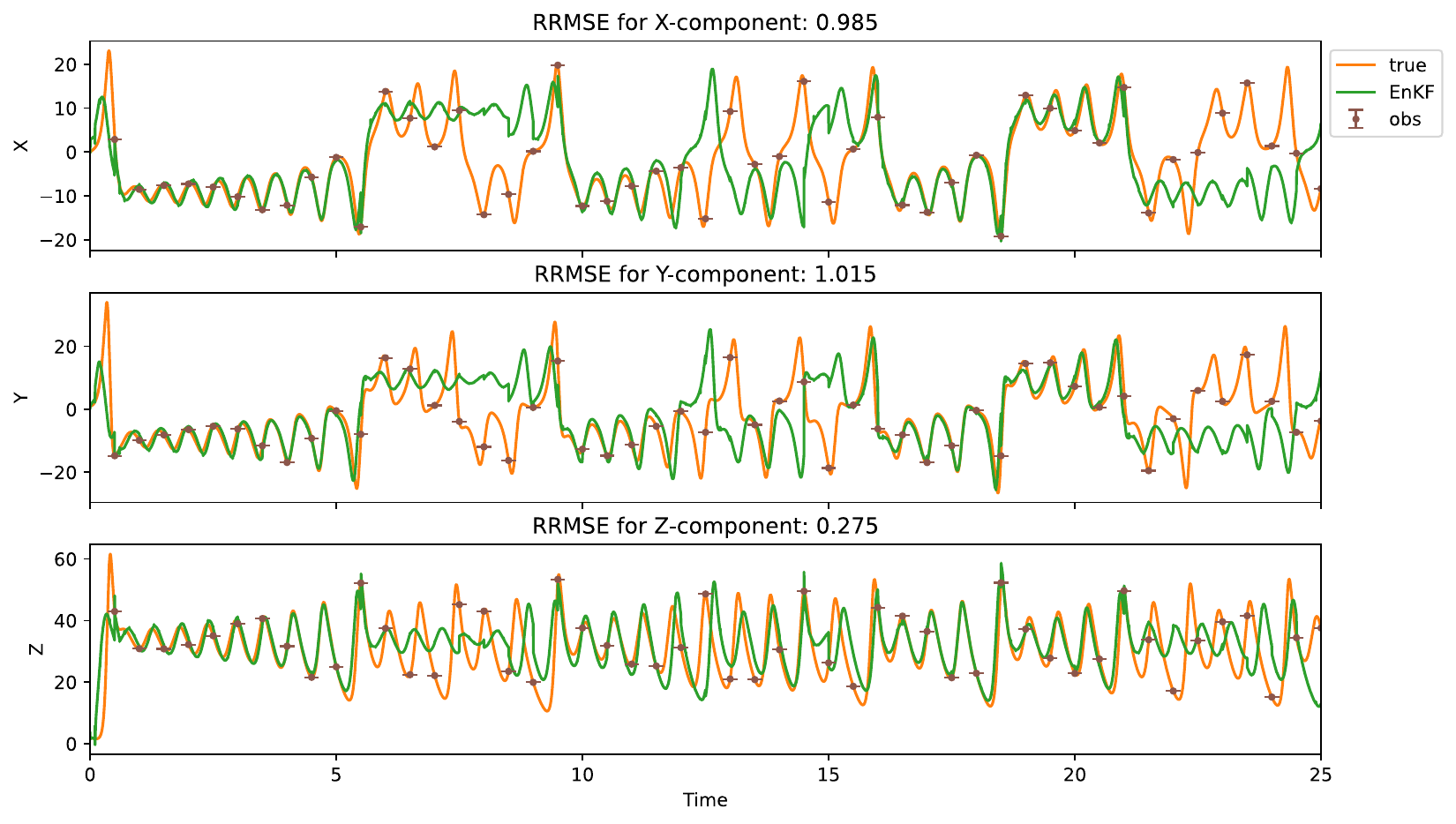}
         \caption{\label{fig:reference_traj_and_enkf}RRMSE on each Component between Reference Trajectory and Trajectory of Forward Model Prediction with EnKF Correction}
     \end{subfigure}
        \caption{RRMSE on each Component between Reference Trajectory and Trajectory of Forward Model Prediction with or without EnKF}
        \label{fig:enkf and without DA comparison graphs}
\end{figure}

In this study, the chosen assimilation algorithm is the EnKF, which comprises 50 ensemble members. The standard deviation of the ensemble members in this case is set to $1$ \blue{for all three components $X, Y, and\ Z$}. Following the computation of the RRMSE between the forward model predictions with EnKF correction and the reference trajectory, Figure \ref{fig:reference_traj_and_enkf} shows the predictions of the forward model post-EnKF correction, alongside the reference trajectory, complete with the RRMSE profiles for each component.

It is noticeable that the EnKF algorithm assisted predictions to achieve \blue{$90.94\% $} and \blue{$69.35\%$} improvements in MSE and RRMSE respectively. \blue{The result shows a promising expectation of engaging EnKF as a foundation algorithm in the data assimilation package, which supports both nonlinear forward model and nonlinear observation operator.}

\subsubsection{Experiment of 4DVar}
\begin{figure}
     \centering
     \begin{subfigure}[t]{1\textwidth}
         \centering
         \includegraphics[width=\textwidth]{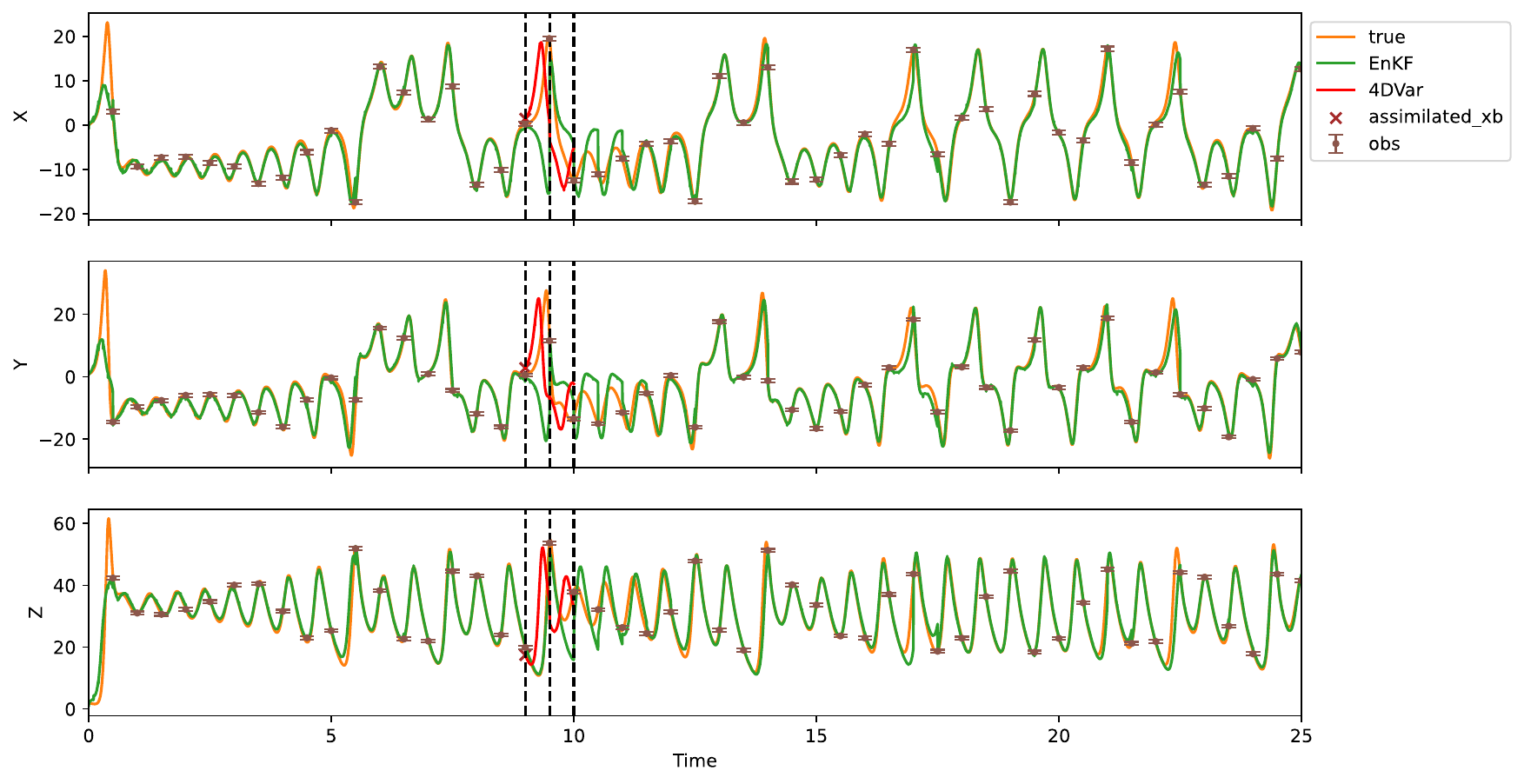}
         \caption{\label{fig:reference_traj_fp_enkf_4dvar_window}\blue{Reference Trajectory, 4DVar Prediction Trajectory with Bounded Vertical Black Dotted Lines, and Trajectory of EnKF Correction on Forward Model Prediction}}
     \end{subfigure}
     \begin{subfigure}[t]{1\textwidth}
         \centering
         \includegraphics[width=\textwidth]{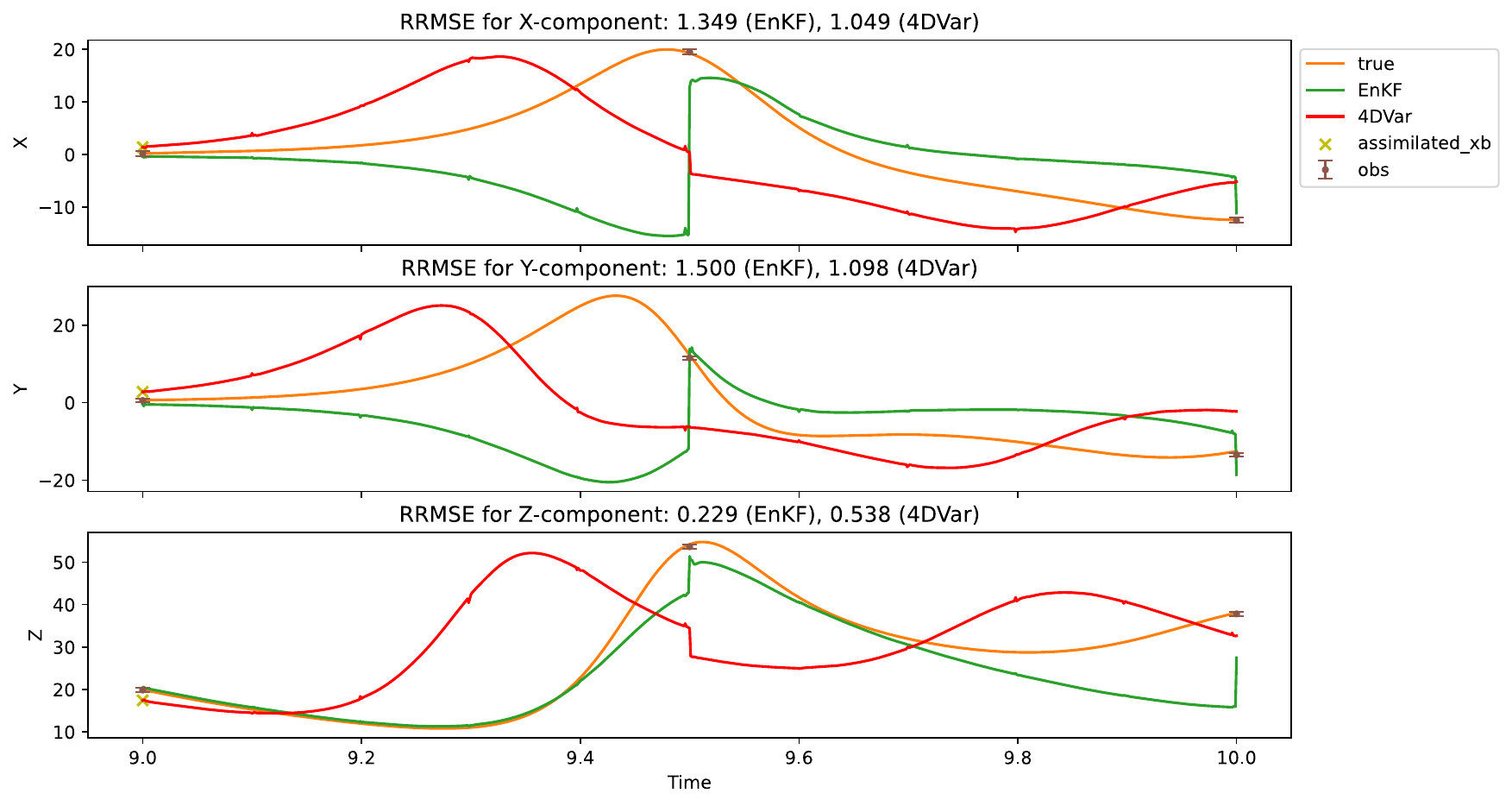} \caption{\label{fig:reference_traj_fp_enkf_4dvar_zoom} \blue{Reference Trajectory, 4DVar Prediction Trajectory, and Trajectory of EnKF Correction on Forward Model Prediction (Zoom View for Time Window Bounded by Vertical Black Dotted Lines in Figure }\ref{fig:reference_traj_fp_enkf_4dvar_window})}
     \end{subfigure}
        \caption{\blue{Reference Trajectory, 4DVar Prediction Trajectory, and Trajectory of EnKF Correction on Forward Model Prediction}}
        \label{fig:4dvar and enkf comparison graphs}
\end{figure}

\blue{In contrast to EnKF, 4DVar directly optimises the evolution trajectory of the system based on all observations and starting background state in the entire evolution trajectory.} In this scenario, three specific observation points indicated by three vertical dotted lines in Figure \ref{fig:reference_traj_fp_enkf_4dvar_window} were selected as constituents of the 4DVar process. \blue{The standard deviation of the background error is set to $1$ for all three components $X, Y, and\ Z$, and the standard deviation of the observation noise was configured to $0.5$. Both standard deviation of the background error and the observation noise are the same configuration as the previous experiment.} Precisely, the optimisation algorithm employed in 4DVar is Adam~\cite{kingma2014adam}, with a specified learning rate of $0.5$. The optimisation process encompasses \blue{200} iterations. \blue{Upon completion of the 4DVar process, it is noticeable from Figure \ref{fig:reference_traj_fp_enkf_4dvar_window} that the 4DVar assimilated result trajectory has a more similar shape to the reference trajectory when compared to the EnKF trajectory at the same window. The RRMSE analysis was conducted to quantitatively evaluate this performance.} To ensure an equitable comparison between the performance of EnKF and 4DVar, both algorithms were subjected to RRMSE calculations within the time interval demarcated by the vertical black dotted lines. Figure \ref{fig:reference_traj_fp_enkf_4dvar_zoom} provides a closer examination of the prediction trajectories generated by both EnKF and 4DVar in comparison to the reference trajectory. Additionally, this figure includes the RRMSE values for both EnKF and 4DVar prediction trajectories. Meanwhile, for explicit demonstration and comparative purposes, all evaluation metrics are presented in Table \ref{tab:lorenz table}.

\blue{It is discernible that the prediction trajectory achieved through the 4DVar and the EnKF algorithm achieved similar quantitative results. The entire evolutionary process within the specified time window of 4DVar assimilated result achieved slightly smoother trajectory compared with the EnKF result. Nevertheless, it is important to acknowledge that the entire qualitative result indicates the EnKF result achieved better result in this assimilation window. This is because EnKF has the property of dynamic representation of errors, which causes it to surpass 4DVar in terms of distance measurement to the reference trajectory~\cite{bocquet2013joint}.}

\begin{table}
    \centering
    \caption{Evaluation Metrics in Different Time Window for the Lorenz Example}
    \label{tab:lorenz table}
    \begin{tabular}{llcccc}
        \hline
        & & 0 to 25 \blue{units} & & 7 to 8 \blue{units} & \\
        \hline
        No Assimilation & MSE & \blue{158.31} & & \blue{241.47} & \\
        & RRMSE & \blue{0.62} & & \blue{0.79} & \\
        \hline
        EnKF & MSE & \blue{14.35} & \blue{(90.94\%$\uparrow$)} & \blue{177.46} & \blue{(26.51\%$\uparrow$)} \\
        & RRMSE & \blue{0.19} & \blue{(69.35\%$\uparrow$)} & \blue{0.68} & \blue{(13.92\%$\uparrow$)} \\
        \hline
        4DVar & MSE & - & & \blue{179.50} & \blue{(25.66\%$\uparrow$)} \\
        & RRMSE & - & & \blue{0.68} & \blue{(13.92\%$\uparrow$)} \\
        \hline
    \end{tabular}
\end{table}

\subsection{Shallow Water Model}
A standard shallow-water fluid mechanics system is commonly employed as a benchmark for assessing the performance of data assimilation algorithms, including 3DVar and 4DVar~\cite{stewart2008correlated,cioaca2014low}. This system represents a nonlinear and time-dependent wave propagation problem. In this scenario, the initial condition is established as a cylindrical body of water with a specified radius, released at $t=0$. It is assumed that the horizontal length scale on the two-dimensional surface holds greater significance than the vertical scale perpendicular to the surface, and the Coriolis force is intentionally disregarded. These factors give rise to the Saint-Venant equations~\cite{venant71Theorie}, which couple velocity components of the fluid (represented as $u$ and $v$) in two dimensions, measured in $0.1 m/s $, with the height of the fluid ($h$), expressed in millimetres. These equations can be formally expressed as equations from Eq.4 to 8. It is noteworthy that in this context, the vertical direction is denoted as the downward $y$ direction. The gravitational constant ($g$), representative of the gravity of Earth, is appropriately scaled to unity ($1$), and the dynamical system is formulated in a non-conservative form.

\begin{align}
& \frac{\partial u}{\partial t} = -g \frac{\partial h}{\partial x}-b u \\
& \frac{\partial v}{\partial t} = -g \frac{\partial h}{\partial y}-b v \\
& \frac{\partial h}{\partial t} = -\frac{\partial(u h)}{\partial x}-\frac{\partial(v h)}{\partial y} \\
& u_{t=0} = 0 \\
& v_{t=0} = 0
\end{align}

\subsubsection{Experimental Settings}
\begin{figure}
    \begin{center}
        \includegraphics[width=\textwidth]{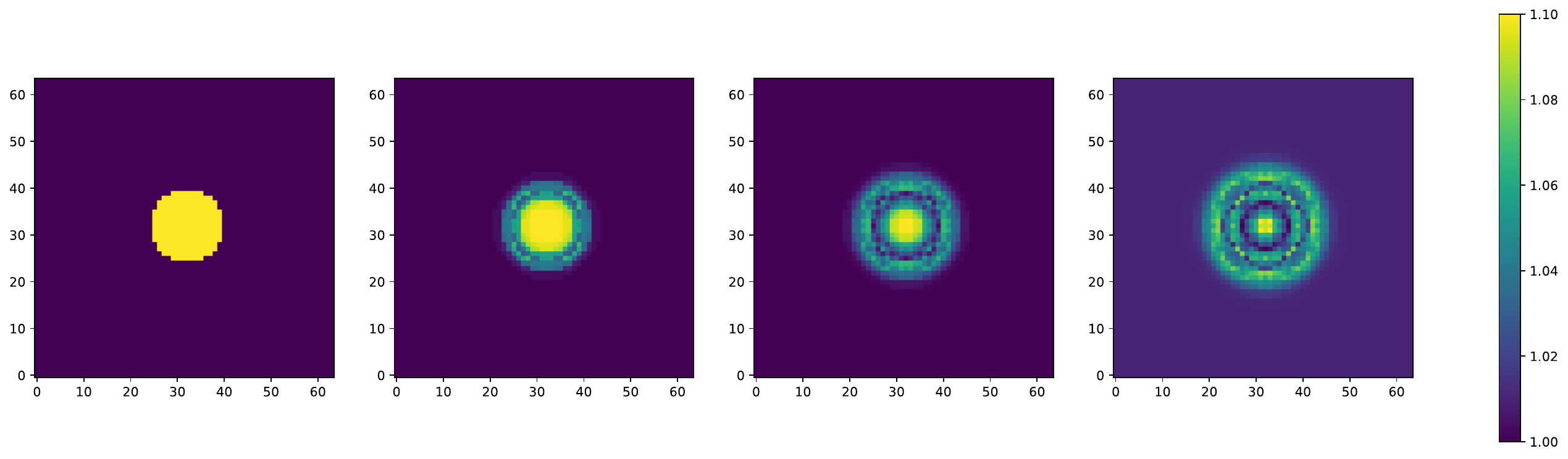}
    \end{center}
    \caption{\label{fig:wave_propagation}A Sequence of Successive Wave Propagation Snapshots of the Water Cylinder Height Start from Initial Condition at $k=0$ to $k=0.06s$ with Time Interval $\Delta k = 200 \times \delta k = 0.02s$}
\end{figure}

The initial values of u and v are both set to zero for the entire velocity field, and the height of the water cylinder is positioned at a height of $0.1mm$ with $8mm$ radius above the still water at the central location of the simulation area. The domain size $(L_x \times L_y)$ is $(64mm \times 64mm)$, discretised with a regular square grid of size $(64 \times 64)$. Equations (4 to 8) are approximated using a first-order finite difference method, and discrete time $k$ replaced the continuous time $t$. Time integration is also first order, with a time interval of $\delta k = 10^{-4}s$, extending the simulation up to $k_\text{final} = 0.8s$. To provide a visual demonstration of the initial condition and the subsequent wave propagation pattern mentioned earlier, a sequence of successive snapshots illustrating the height of the water cylinder is presented in Figure~\ref{fig:wave_propagation}.

At each evolution step, the entire simulation field for physical quantities u ($\mathbf{x}_k$) and h ($\mathbf{y}_k$) is recorded as grayscale images at full size $(64 \times 64)$. These images of u are treated as the full state space variable $\{\mathbf{x}_k\}$. A convolutional autoencoder is trained on these $\mathbf{x}_k$ images to compress them from their original $(64 \times 64)$ size to a 1D latent space vector representation $ \hat{\mathbf{x}}_k$ with 32 elements, which is then reconstructed back to the original image. Another convolutional autoencoder with the same architecture is trained on $\mathbf{y}_k$ using the same process as the autoencoder for $\mathbf{x}_k$. These encoders trained on each autoencoder are utilised in the subsequent latent data assimilation process. Additionally, a convolutional deep residual network is trained to map a given $\mathbf{x}_k$ image to its corresponding $\mathbf{y}_k$ image at the same time instance. This deep residual network serves as a model for the observation operator $\mathcal{H}$, which maps full space $\mathbf{x}_k$ to full space $\mathbf{y}_k$. These statements can also be denoted by mathematical expressions presented in Equations 9 to 13. There was 1\% randomly selected data (80 samples) in all generated data utilised in validation for $\mathbf{x}_k$ autoencoder, $\mathbf{y}_k$ autoencoder, and observation operator $\mathcal{H}$. All remaining data were adopted for training these neural networks.

\begin{align}
& \text{state}\ \mathbf{x}_k = D_u(E_u(\mathbf{x}_k)) \gets u \\
& \text{latent state}\ \hat{\mathbf{x}}_k = E_u(\mathbf{x}_k) \\
& \text{observation}\ \mathbf{y}_k = D_h(E_h(\mathbf{y}_k)) = \mathcal{H}(\mathbf{x}_k) = \mathcal{H}(D_u(\hat{\mathbf{x}}_k)) \gets h \\
& \text{observation operator}\ \hat{\mathcal{H}}(\hat{\mathbf{x}}_k) = E_h(\mathcal{H}(D_u(\hat{\mathbf{x}}_k))) \\
& \text{latent observation}\ \hat{\mathbf{y}}_k = \hat{\mathcal{H}}(\hat{\mathbf{x}}_k) = E_h(\mathbf{y}_k) \\
& \mathbf{B} = \mathbf{I}_{(32\times 32)} \quad \textrm{(the latent space is of dimension 32)}\\
& \mathbf{R}_{\text{full}} = 10^{-10} \cdot \mathbf{I}_{(64\times 64)^2} \textrm{(the full space is of dimension $64 \times 64$)} \\
& \mathbf{R}_{\text{latent}} = 10^{-10} \cdot \mathbf{I}_{(32\times 32)}
\end{align}

Moreover, a 20-layer LSTM network is employed to learn the evolution of $\mathbf{x}_k$ in latent space, with an input dimension of 32 and a hidden dimension of 256. This LSTM network was validated on 1\% sequential data (80 samples) selected at the end of the whole simulation window, and all other data were utilised in training the network. Following all processes above, the only missing component is the operator $\hat{\mathcal{H}}$, which can perform reduced space assimilation. This operator can be constructed by combining various elements from previous works, including 1) the decoder of $\mathbf{x}_k$ from the $\mathbf{x}_k$ autoencoder, 2) the model $\mathcal{H}$ mapping full space $\mathbf{x}_k$ to full space $\mathbf{y}_k$, and 3) the encoder of $\mathbf{y}_k$ from the $\mathbf{y}_k$ autoencoder, arranged in a sequential order.

\subsubsection{3DVar: Latent Space \texorpdfstring{$\hat{\mathbf{x}}_k$}{hxk} to Full Space \texorpdfstring{$\mathbf{y}_k$}{yk}}
\label{3DVar: Latent Space u to Full Space h}
In this scenario, the latent assimilation was conducted by using an observation operator $\mathcal{H}(D_u(\hat{\mathbf{x}}_k))$ capable of mapping the latent space $\hat{\mathbf{x}}_k$ to the full space $\mathbf{y}_k$. This operator is constructed using the decoder from the autoencoder $\mathbf{x}_k$ and the full space $\mathbf{x}_k$ to full space $\mathbf{y}_k$ observation operator denoted as $\mathcal{H}$. In the case of 3DVar, there are three selected specific time instances for latent assimilation. To be precise, latent assimilation is performed every 50 forward predictions, corresponding to the 1850th ($t_1=0.185s$), 1900th ($t_2=0.19s$), and 1950th ($t_3=0.195s$) recording steps within the full simulation time window.

During each assimilation time point, the cost function for optimisation is constructed using $\mathbf{B}$ in Eq. 14 and $\mathbf{R}_{full}$ in Eq. 15. The optimisation algorithm employed remains Adam~\cite{kingma2014adam}, with a learning rate of 5 and a maximum of 300 iterations.

\begin{figure}
     \centering
     \begin{subfigure}[c]{0.45\textwidth}
         \centering
         \includegraphics[width=\textwidth]{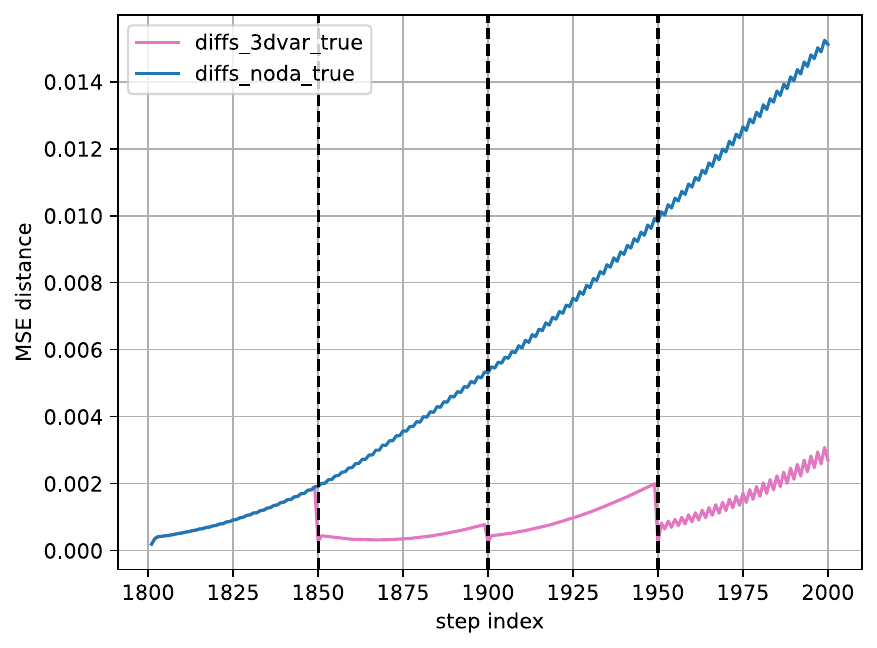}
         \caption{\label{fig:3dvar_lu2fh}MSE for Latent Space $\hat{\mathbf{x}}_k$ to Full Space $\mathbf{y}_k$ 3DVar Latent Assimilation}
     \end{subfigure}
     \begin{subfigure}[c]{0.45\textwidth}
         \centering
         \includegraphics[width=\textwidth]{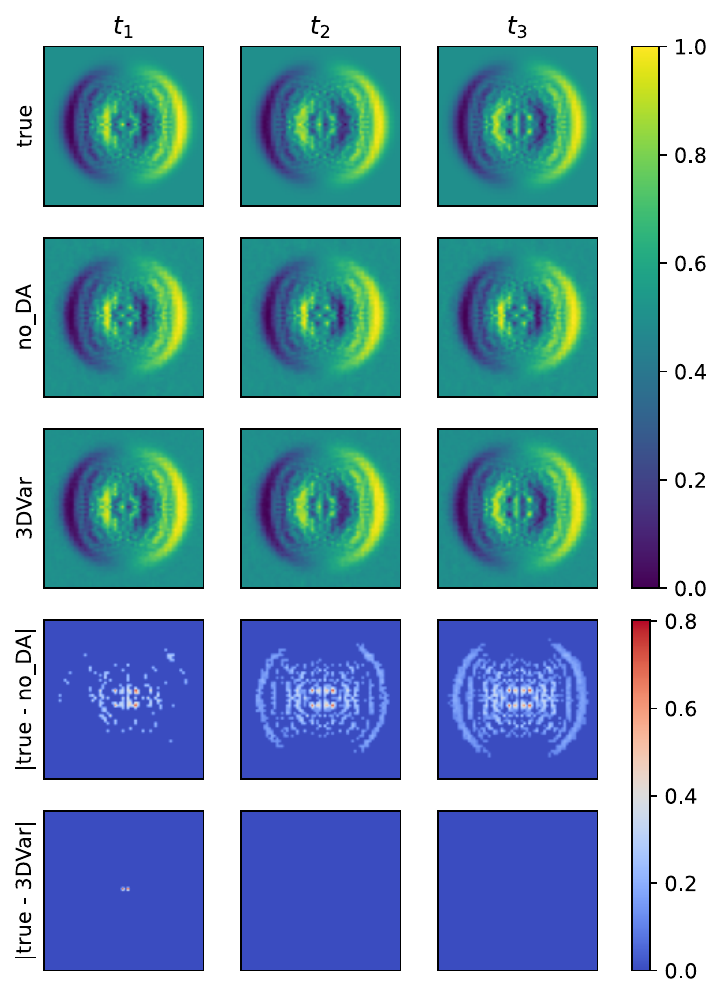}
         \caption{\label{fig:3dvar_lu2fh_imgs}Latent Space $\hat{\mathbf{x}}_k$ to Full Space $\mathbf{y}_k$ 3DVar Images Showcase (Absolute Differences below 0.1 Truncated)}
     \end{subfigure}
        \caption{Latent Space $\hat{\mathbf{x}}_k$ to Full Space $\mathbf{y}_k$: MSE for 3DVar Latent Assimilation and Images Showcase}
        \label{fig:3dvar_lu2fh and 3dvar_lu2fh_imgs}
\end{figure}

At each step of the latent vector prediction, decoding is performed using the decoder of the autoencoder $\mathbf{x}_k$. Figure~\ref{fig:3dvar_lu2fh_imgs} presents various elements, comprising the original $\mathbf{x}_k$ image, the decoded prediction image, the decoded image derived from 3DVar assimilation, the absolute difference between the original $\mathbf{x}_k$ image and the decoded prediction image (with differences below 0.1 truncated), and the absolute difference between the original $\mathbf{x}_k$ image and the decoded image produced through 3DVar assimilation (with differences below 0.1 truncated). Subsequently, the Mean Squared Error (MSE) is calculated between the reconstructed image and the original image recorded at the corresponding time instance. The resulting curve depicting the evolution of MSE over time steps is presented as Figure~\ref{fig:3dvar_lu2fh}. By observing both evaluation metric plot and images showcase, it is noticeable in Figure~\ref{fig:3dvar_lu2fh and 3dvar_lu2fh_imgs} that each application of 3DVar is capable of reducing model prediction errors by obtaining a better state estimate at the observation step.

\subsubsection{3DVar: Latent Space \texorpdfstring{$\hat{\mathbf{x}}_k$}{hxk} to Latent Space \texorpdfstring{$\hat{\mathbf{y}}_k$}{hyk}}
\begin{figure}
     \centering
     \begin{subfigure}[c]{0.45\textwidth}
         \centering
         \includegraphics[width=\textwidth]{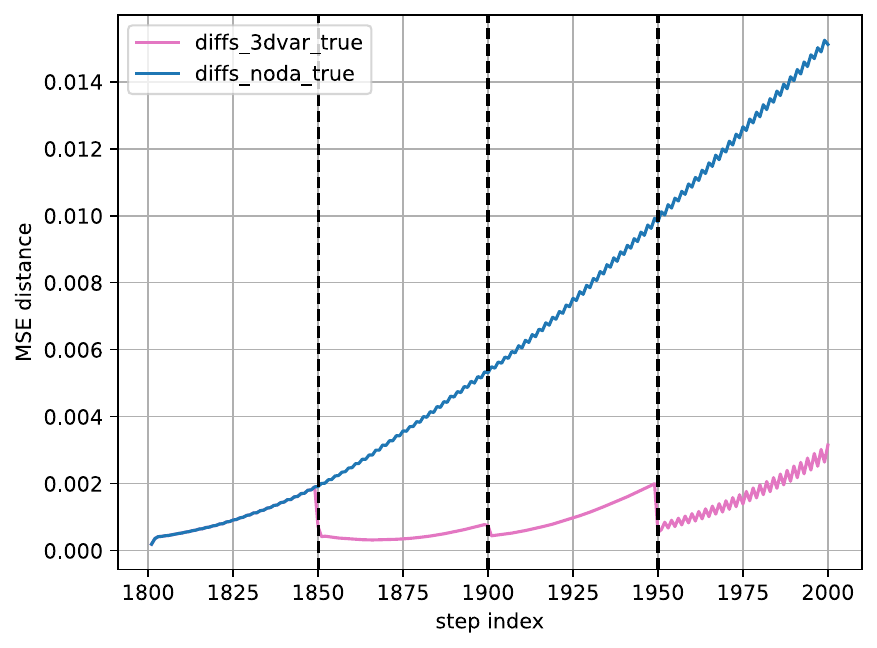}
         \caption{\label{fig:3dvar_lu2lh}MSE for Latent Space $\hat{\mathbf{x}}_k$ to Latent Space $\hat{\mathbf{y}}_k$ 3DVar Latent Assimilation}
     \end{subfigure}
     \begin{subfigure}[c]{0.45\textwidth}
         \centering
         \includegraphics[width=\textwidth]{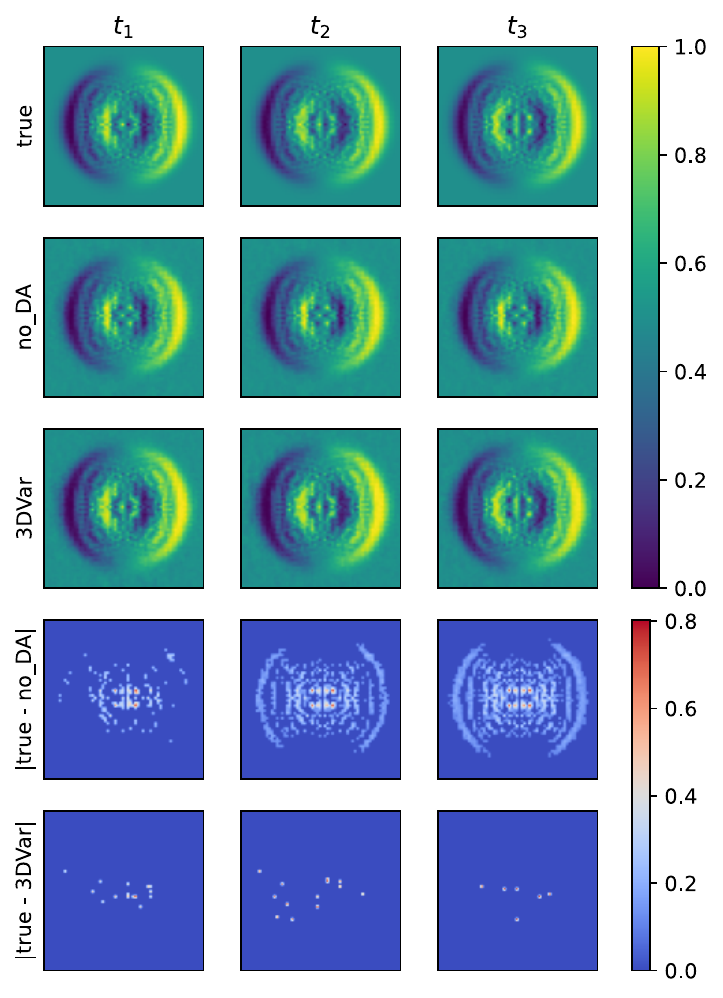}
         \caption{\label{fig:3dvar_lu2lh_imgs}Latent Space $\hat{\mathbf{x}}_k$ to Latent Space $\hat{\mathbf{y}}_k$ 3DVar Images Showcase (Absolute Differences below 0.1 Truncated)}
     \end{subfigure}
        \caption{Latent Space $\hat{\mathbf{x}}_k$ to Latent Space $\hat{\mathbf{y}}_k$: MSE for 3DVar Latent Assimilation and Images Showcase}
        \label{fig:3dvar_lu2lh and 3dvar_lu2lh_imgs}
\end{figure}

In this context, latent assimilation is conducted, utilising the observation operator denoted as $\hat{\mathcal{H}}$, responsible for mapping the latent space $\hat{\mathbf{x}}_k$ to the latent space $\hat{\mathbf{y}}_k$. $\hat{\mathcal{H}}$ is a MLP neural network, trained separately to link the latent state space and the latent observation space (see Eq. 13). Following the previous approach, the same three time instances were selected for latent assimilation. At each assimilation time point, the optimisation process constructs a cost function utilising $\mathbf{B}$ in Eq. 14 and $\mathbf{R}_\text{latent}$ in Eq. 16. The settings of the optimisation algorithm remain unchanged, and evaluation procedure also remains the same as before. To provide a comprehensive assessment for the performance of the latent assimilation, the RRMSE and the Structural Similarity Index Measure (SSIM) were evaluated, in addition to the MSE metric. These evaluation metrics are presented in Table~\ref{tab:3dvar table}. As shown by Figure~\ref{fig:3dvar_lu2lh and 3dvar_lu2lh_imgs}, it is evident that both latent space $\hat{\mathbf{x}}_k$ to full space $\mathbf{y}_k$ and latent space $\hat{\mathbf{x}}_k$ to latent space $\hat{\mathbf{y}}_k$ assimilation are effectively reducing model prediction errors, but these reconstructed images presented that the assimilation involved in latent space $\hat{\mathbf{x}}_k$ to full space $\mathbf{y}_k$ slightly outperforms the latent data assimilation scheme with $\hat{\mathbf{x}}_k$ and $\hat{\mathbf{y}_k}$. One possible reason is that full space $\mathbf{y}_k$ contains more effective information than highly compressed latent space $\hat{\mathbf{y}}_k$, but the assimilation scheme adopted in Section~\ref{3DVar: Latent Space u to Full Space h} is relatively computational expensive. This is because the optimisation step contains the calculation of a noticeably larger observation covariance matrix $\mathbf{R}$.

\begin{table}
    \centering
    \caption{Evaluation Metrics on Each Data Assimilation (DA) Point in 3DVar}
    \label{tab:3dvar table}
    \begin{tabular}{llcccccc}
        \hline
        time step &  & 1850th & & 1900th & & 1950th & \\
        \hline
        No Assimilation & MSE & 0.0019 & & 0.0053 & & 0.0098 & \\
        & RRMSE & 0.0830 & & 0.1382 & & 0.1881 & \\
        & SSIM & 0.8974 & & 0.7700 & & 0.6080 & \\
        \hline
        Latent Space $\hat{\mathbf{x}}_k$ to & MSE & 0.0003 & (84.21\%$\uparrow$) & 0.0003 & (94.34\%$\uparrow$) & 0.0002 & (97.96\%$\uparrow$)\\
        Full Space $\mathbf{y}_k$ & RRMSE & 0.0316 & (61.93\%$\uparrow$) & 0.0308 & (77.71\%$\uparrow$) & 0.0290 & (84.58\%$\uparrow$)\\
        & SSIM & 0.9726 & (7.73\%$\uparrow$) & 0.9707 & (20.68\%$\uparrow$) & 0.9745 & (37.61\%$\uparrow$)\\
        \hline
        Latent Space $\hat{\mathbf{x}}_k$ to & MSE & 0.0007 & (63.16\%$\uparrow$) & 0.0007 & (86.79\%$\uparrow$) & 0.0005 & (94.90\%$\uparrow$)\\
        Latent Space $\hat{\mathbf{y}}_k$ & RRMSE & 0.0507 & (38.92\%$\uparrow$) & 0.0518 & (62.52\%$\uparrow$) & 0.0443 & (76.45\%$\uparrow$)\\
        & SSIM & 0.9473 & (5.27\%$\uparrow$) & 0.9459 & (18.60\%$\uparrow$) & 0.9572 & (36.48\%$\uparrow$) \\
        \hline
    \end{tabular}
\end{table}

\subsubsection{4DVar: Latent Space \texorpdfstring{$\hat{\mathbf{x}}_k$}{hxk} to Full Space \texorpdfstring{$\mathbf{y}_k$}{yk}}
\label{4DVar: Latent Space u to Full Space h}
\begin{figure}
     \centering
     \begin{subfigure}[c]{0.45\textwidth}
         \centering
         \includegraphics[width=\textwidth]{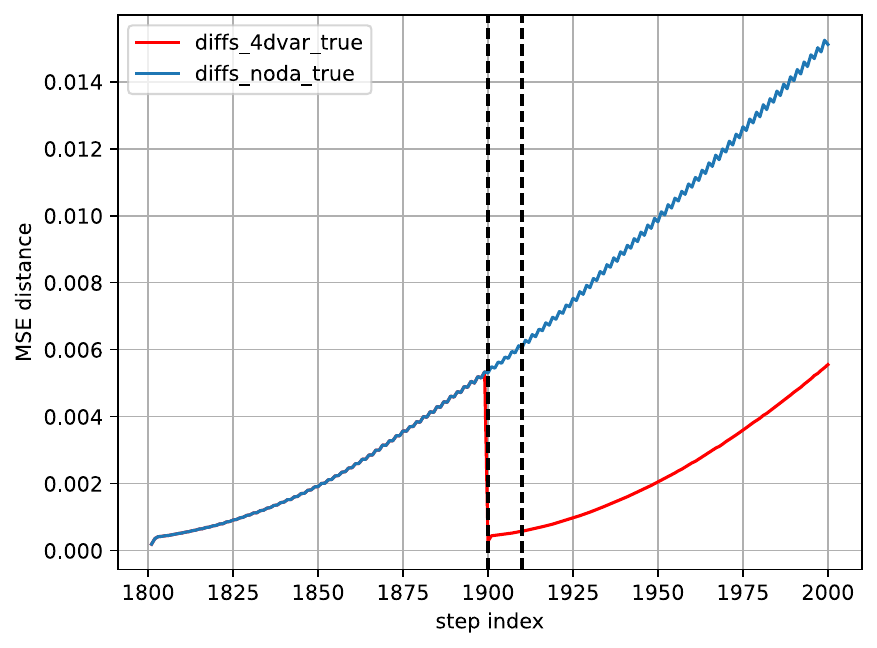}
         \caption{\label{fig:4dvar_lu2fh}MSE for Latent Space $\hat{\mathbf{x}}_k$ to Full Space $\mathbf{y}_k$ 4DVar Latent Assimilation (Assimilation Window: 1900 to 1910)}
     \end{subfigure}
     \begin{subfigure}[c]{0.3\textwidth}
         \centering
         \includegraphics[width=\textwidth]{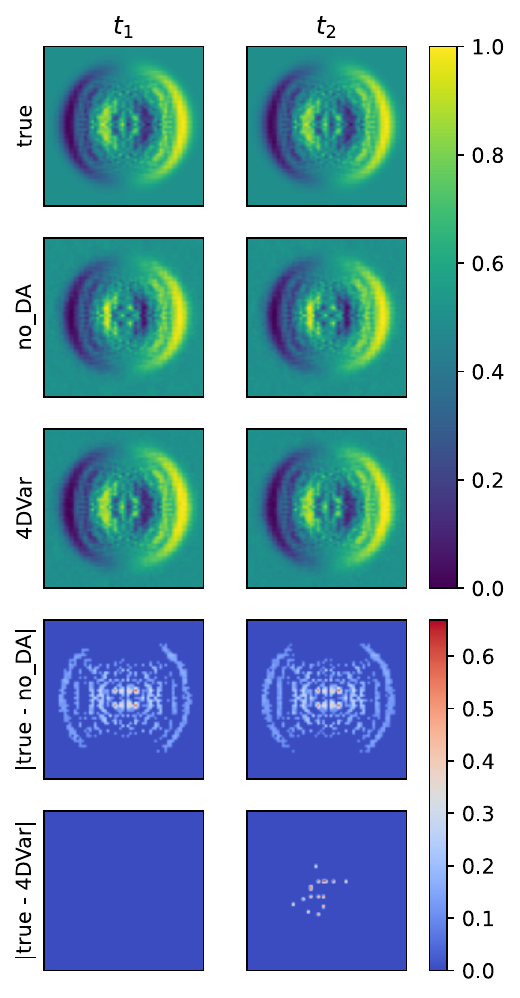}
         \caption{\label{fig:4dvar_lu2fh_imgs}Latent Space $\hat{\mathbf{x}}_k$ to Full Space $\mathbf{y}_k$ 4DVar Images Showcase (Absolute Differences below 0.1 Truncated)}
     \end{subfigure}
        \caption{Latent Space $\hat{\mathbf{x}}_k$ to Full Space $\mathbf{y}_k$: MSE for 4DVar Latent Assimilation and Images Showcase}
        \label{fig:4dvar_lu2fh and 4dvar_lu2fh_imgs}
\end{figure}

In this scenario, latent assimilation is conducted using the full space operator $\mathcal{H}$, consistent with Section~\ref{3DVar: Latent Space u to Full Space h}. The latent assimilation is initiated at the 1900th recorded step ($t_1=0.19s$) and is combined with another observation recorded at the 1910th step ($t_2=0.191s$) within the full simulation time window. At the assimilation time point, the cost function for optimisation is constructed using $\mathbf{B}$ in Eq. 14 and $\mathbf{R}_\text{full}$ in Eq. 15. The optimisation algorithm remains Adam~\cite{kingma2014adam}, with a learning rate of 2.5 and a maximum of 100 iterations. The evaluation procedure remains the same as before. It is noticeable in Figure~\ref{fig:4dvar_lu2fh and 4dvar_lu2fh_imgs} that 4DVar is also capable of fixing model prediction errors by obtaining a better state estimate, but this better estimate involves consideration of all observations in the whole assimilation window.

\subsubsection{4DVar: Latent Space \texorpdfstring{$\hat{\mathbf{x}}_k$}{hxk} to Latent Space \texorpdfstring{$\hat{\mathbf{y}}_k$}{hyk}}
\begin{figure}
     \centering
     \begin{subfigure}[c]{0.45\textwidth}
         \centering
         \includegraphics[width=\textwidth]{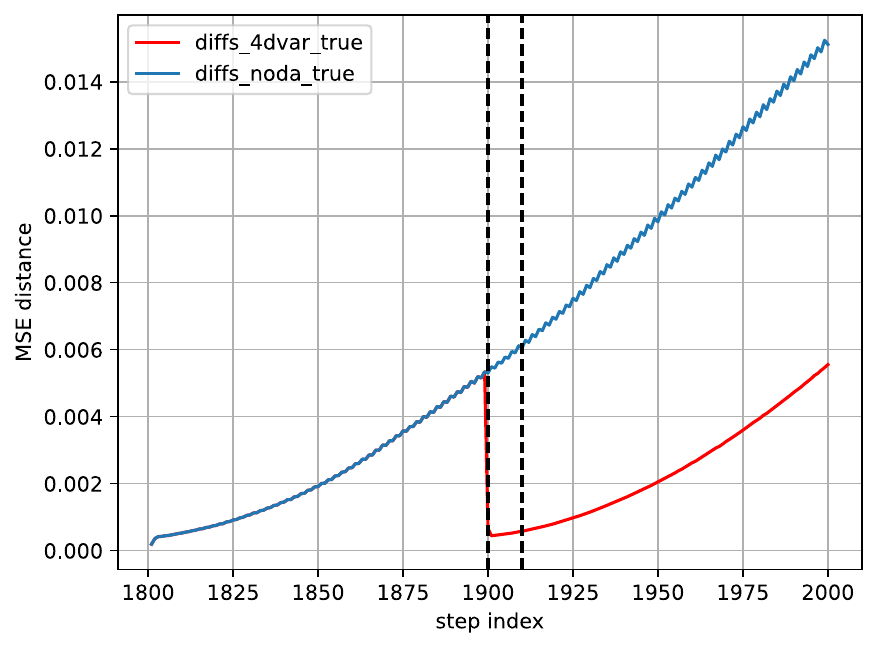}
         \caption{\label{fig:4dvar_lu2lh}MSE for Latent Space $\hat{\mathbf{x}}_k$ to Latent Space $\hat{\mathbf{y}}_k$ 4DVar Latent Assimilation (Assimilation Window: 1900 to 1910)}
     \end{subfigure}
     \begin{subfigure}[c]{0.3\textwidth}
         \centering
         \includegraphics[width=\textwidth]{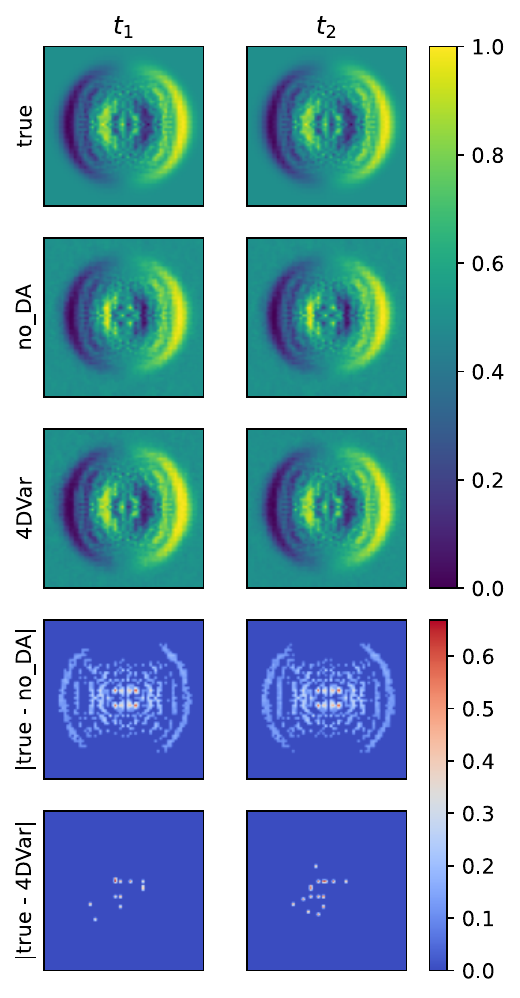}
         \caption{\label{fig:4dvar_lu2lh_imgs}Latent Space $\hat{\mathbf{x}}_k$ to Latent Space $\hat{\mathbf{y}}_k$ 4DVar Images Showcase (Absolute Differences below 0.1 Truncated)}
     \end{subfigure}
        \caption{Latent Space $\hat{\mathbf{x}}_k$ to Latent Space $\hat{\mathbf{y}}_k$: MSE for 4DVar Latent Assimilation and Images Showcase}
        \label{fig:4dvar_lu2lh and 4dvar_lu2lh_imgs}
\end{figure}

In this scenario, latent assimilation is conducted utilising the observation operator denoted as $\hat{\mathcal{H}}$. Similar to the previous approach, all the same time instances were selected as the assimilation points. At these designated assimilation points, the cost function for optimisation is constructed utilising $\mathbf{B}$ in Eq. 14 and $\mathbf{R}_\text{latent}$ in Eq. 16. The settings of the optimisation algorithm remains unchanged, and the evaluation procedure remains the same as before. To further inspect the performance of the latent assimilation results, assessments using the RRMSE and the SSIM in addition to the MSE metric were also conducted, and these findings are presented in Table~\ref{tab:4dvar table} and Figure~\ref{fig:4dvar_lu2lh and 4dvar_lu2lh_imgs}. It is also evident that both assimilation schemes involving 4DVar can effectively reduce model prediction errors, and the scheme contains full space $\mathbf{y}_k$ still slightly outperforms another scheme stated in this section. However, similar to the 3Dvar case, the computational cost for the assimilation scheme in Section~\ref{4DVar: Latent Space u to Full Space h} is also relatively expensive due to the optimisation step including a significantly larger matrix $\mathbf{R}$.

\begin{table}
    \centering
    \caption{Evaluation Metrics on Each Data Assimilation (DA) Point in 4DVar}
    \label{tab:4dvar table}
    \begin{tabular}{llccccc}
        \hline
        & & 1900th & & 1910th & \\
        \hline
        No Assimilation & MSE & 0.0053 & & 0.0060 & \\
        & RRMSE & 0.1382 & & 0.1477 & \\
        & SSIM & 0.7700 & & 0.7408 & \\
        \hline
        Latent Space $\hat{\mathbf{x}}_k$ to Full Space $\mathbf{y}_k$ & MSE & 0.0003 & (94.34\%$\uparrow$) & 0.0006 & (90.00\%$\uparrow$)\\
        & RRMSE & 0.0318 & (76.99\%$\uparrow$) & 0.0456 & (69.13\%$\uparrow$)\\
        & SSIM & 0.9697 & (20.59\%$\uparrow$) & 0.9564 & (22.54\%$\uparrow$)\\
        \hline
        Latent Space $\hat{\mathbf{x}}_k$ to Latent Space $\hat{\mathbf{y}}_k$ & MSE & 0.0007 & (86.79\%$\uparrow$) & 0.0006 & (90.00\%$\uparrow$)\\
        & RRMSE & 0.0488 & (64.69\%$\uparrow$) & 0.0456 & (69.13\%$\uparrow$)\\
        & SSIM & 0.9469 & (18.68\%$\uparrow$) & 0.9564 & (22.54\%$\uparrow$)\\
        \hline
    \end{tabular}
\end{table}

\subsubsection{Results and Discussion}
It is evident that all outcomes resulting from the data assimilation process surpass the performance of pure forward predictions conducted without data assimilation. Furthermore, it is notable that, within the entire simulation time window, 3DVar exhibits a slightly superior outcome than 4DVar at the 1900th step for the latent $\hat{\mathbf{x}}_k$ to full space $\mathbf{y}_k$ case. An important contributing factor to this distinction is the difference in optimisation steps, with 3DVar employing 300 iterations, while 4DVar utilises only 100 iterations. This difference arises from 3DVar involving only a one-to-one state assimilation without the necessity of forward predictions, as is required in 4DVar. Consequently, 4DVar is relatively computationally expensive. In reality, the computational time for 300 iterations in 3DVar and 100 iterations in 4DVar is roughly equivalent in benchmark testing. Hence, it becomes important for users to carefully balance computational costs against the quality of results when selecting an algorithm. In addition, the average inference time for all experiments conducted in the shallow water model has been listed in Table~\ref{tab:benchmarking table}, and all tests were conducted on a NVIDIA GeForce RTX 3080 Ti Laptop GPU. The benchmarking results indicate the relatively acceptable time consumption for data assimilation aided forward predictions.

\begin{table}
    \centering
    \caption{Average Inference Time for the 20-layer LSTM from 1800th Record to 2000th Record}
    \label{tab:benchmarking table}
    \begin{tabular}{llcc}
        \hline
        & & No Assimilation & With Assimilation \\
        \hline
        3DVar & Latent Space \texorpdfstring{$\hat{\mathbf{x}}_k$}{hxk} to Full Space \texorpdfstring{$\mathbf{y}_k$}{yk} & 0.01s & 0.43s \\
        & Latent Space \texorpdfstring{$\hat{\mathbf{x}}_k$}{hxk} to Latent Space \texorpdfstring{$\hat{\mathbf{y}}_k$}{hyk} & 0.01s & 0.07s \\
        \hline
        4DVar & Latent Space \texorpdfstring{$\hat{\mathbf{x}}_k$}{hxk} to Full Space \texorpdfstring{$\mathbf{y}_k$}{yk} & 0.01s & 0.13s \\
        & Latent Space \texorpdfstring{$\hat{\mathbf{x}}_k$}{hxk} to Latent Space \texorpdfstring{$\hat{\mathbf{y}}_k$}{hyk} & 0.01s & 0.05s \\
        \hline
    \end{tabular}
\end{table}

By testing EnKF, 3DVar, and 4DVar that are commonly adopted in data assimilation on Lorenz 63 system and shallow water example, the result on each test case indicates that the package was implemented correctly. In the whole process of utilising the package, it only requires users to directly setup relevant parameters in a CaseBuilder object before launch the execution of the actual algorithm. $\mathcal{M}_k$ and $\mathcal{H}$ can be set at any time before running the algorithm, and both should be callable objects, which is a loose constrain on $\mathcal{M}_k$ and $\mathcal{H}$. This package is capable of handling either explicit or implicit $\mathcal{M}_k$ and $\mathcal{H}$ in any neural network architectures, which enhanced flexibility and compatibility. Similarly, the data assimilation algorithm is also provided as an option inside, and it can also be set at any time before launch the execution. Users have at least three different approaches to setup parameters such as algorithms, $\mathcal{M}_k$, and $\mathcal{H}$, which contains set by Parameters object, dictionary passing, and setter methods calling. All setter methods support chained calling, so users are less likely to setup parameters incorrectly. Therefore, this package provides user friendly interfaces and not loss flexibility of handling nonlinear state transformation function $\mathcal{M}_k$ and observation operator $\mathcal{H}$ with any neural network architectures.

\section{Conclusion and future work}
In this study, a novel Python package TorchDA was developed for seamlessly integrating data assimilation with deep learning models. TorchDA provides implementations of various algorithms including Kalman Filter, EnKF, 3DVar, and 4DVar methods. It allows users to substitute conventional numerical models with neural networks trained as models for the state transformation function and observation operator.
This novel package offers the capability to conduct variational or Kalman-type data assimilation using a non-explicit transformation function, which is represented by neural networks. This feature addresses a significant limitation of current data assimilation tools, which are unable to handle such non-explicit functions.

The functionality and effectiveness of the package were demonstrated through comprehensive experiments on the Lorenz 63 system and shallow water equations. The Lorenz 63 example illustrated the capabilities of EnKF and 4DVar algorithms integrated with multilayer LSTM networks to track the reference trajectory. The shallow water model assessment involved autoencoders and residual networks acting as observation operators between different physical quantities, validating the latent assimilation workflow. In both experiments, assimilated results using the package significantly outperformed raw model predictions without assimilation. Overall, this innovative package offers researchers a flexible tool to harness the representation power of deep learning for modelling within the data assimilation paradigm, including some examples such as ocean climate prediction, nuclear engineering, hydraulics, and fluid dynamics presented in the introduction section.

Future work can focus on expanding the built-in algorithm collection with more advanced techniques, such as differentiable EnKF~\cite{liu2023enhancing}, sigma-point Kalman Filters (SPKF), and Particle Filter (PF)~\cite{tang2016introduction}. This package can also incorporate hybrid data assimilation techniques in the future, including ETKF-3DVAR~\cite{wang2008hybrid}, Ensemble 4DVar (En4DVar)~\cite{zhu2022four}, and 4D Ensemble-Variational (4DEnVar)~\cite{zhu2022four} methods. Furthermore, it is worth investigating scientific methods for evaluating covariance matrices $\mathbf{B}_k$ and $\mathbf{R}_k$ in the future. The package demonstrates promising potential to facilitate novel deep learning and data assimilation integrated solutions, advancing scientific research across domains characterised by high dimensional nonlinear dynamics. \rev{Effort will also be dedicated to applying the new TorchDA package to real-world data assimilation problems, where more engineering techniques, such as localization and inflation schemes, will be further examined within the proposed package. To reduce the computational burden of 4DVar algorithms, future implementations will include an official setup of the incremental 4DVar algorithms~\cite{farchi2023online} within the TorchDA package. Additionally, for the crucial task of covariance specification in data assimilation, covariance tuning methods (e.g.,~\cite{desroziers2005diagnosis}) are planned for incorporation in future versions.}




\bibliographystyle{abbrv}
\bibliography{references.bib}  

\appendix
\section{Python Interfaces}
\begin{lstlisting}[language=Python, caption=General Structure of the Parameters Class]
@dataclass
class Parameters:
    algorithm
    device
    observation_model
    background_covariance_matrix
    observation_covariance_matrix
    background_state
    observations
"""Following parameters are OPTIONAL"""
    forward_model
    output_sequence_length
    observation_time_steps
    gaps
    num_ensembles
    start_time
    max_iterations
    learning_rate
    record_log
    args
\end{lstlisting}

\begin{lstlisting}[language=Python, caption=General Structure of the CaseBuilder Class]
class CaseBuilder:
    def __init__(self, case_name=None, parameters=None):
        # Initialize case_name and parameters
        # If case_name is None, set it to the current timestamp
        # Initialize parameters as an instance of Parameters class
    # Method to set a batch of parameters
    def set_parameters(self, parameters):
        # If parameters is an instance of Parameters:
        #     Convert it to a dictionary
        # Create a new CaseBuilder object called checked_builder
        # For each parameter in parameters:
        #     Check if the parameter exists in Parameters class
        #     Check if there is a corresponding setter method
        #     Set the parameter using the setter method
        # Update the parameters of this object
    # Method to set an individual parameter
    def set_parameter(self, name, value):
        # Check if the parameter name exists in Parameters class
        # Check if there is a corresponding setter method for the parameter
        # Call the setter method with the given value
    # Methods to set various configuration parameters
    def set_[parameter_name](self, [parameter_name])
    # Method to execute the data assimilation case
    def execute(self):
        # Set input parameters for execution
        # Run the data assimilation case
        # Return the results as a dictionary
    # Methods to retrieve results and parameters
    def get_results_dict(self):
        # Get the dictionary containing results
    def get_result(self, name):
        # Get a specific result by name
    def get_parameters_dict(self):
        # Get the configured parameters as a dictionary
\end{lstlisting}

\begin{lstlisting}[language=Python, caption=General Structure of the \_Executor Class]
class _Executor:
    def set_input_parameters(parameters):
        # Set input parameters for data assimilation
        # Update parameters of the Executor
    def __check_[algorithm]_parameters():
        # Check algorithm parameters for validity
    def __call_apply_[algorithm]():
        # Call apply_[algorithm] with configured parameters
        # Return algorithm results
    def __setup_device():
        # Set up computation device (CPU or GPU) based on user preference
    def run():
        # Run selected data assimilation algorithm
        # Return results as a dictionary
    def get_results_dict():
        # Get deep copy of results dictionary
    def get_result(name):
        # Get deep copy of a specific result by name
\end{lstlisting}

\end{document}